\documentclass{aa}
\usepackage{graphicx}
\usepackage{txfonts}
\begin{document}

\title{Stationary field-aligned MHD flows at astropauses and in astrotails}
\subtitle{Principles of a counterflow configuration between a stellar 
wind and its
interstellar medium wind }

\author{D.H. Nickeler\inst{1,2,3}, J.P. Goedbloed\inst{2,4},
\and
    H.-J. Fahr\inst{3}
}

\offprints{D.H. Nickeler}
\mail{nickeler@asu.cas.cz}

\institute{Astronomical Institute, AV \v{C}R, Fri\v{c}ova 298,
25165 Ond\v{r}ejov, Czech
Republic
\and
Astronomical Institute, University of Utrecht, Princetonplein 5, 3584 
CC Utrecht,
The
                  Netherlands
            \and
            Institute for Astrophysics and Extraterrestrial Research, Auf dem
H\"{u}gel 71,
53121 Bonn, Germany
            \and
            FOM-Institute for Plasma Physics, Edisonbaan 14, 3439 MN Nieuwegein,
the Netherlands
       }

\date{Received; accepted}

\abstract
{A stellar wind
passing through the reverse shock is deflected into the
astrospheric tail and leaves the stellar system either as a 
sub-Alfv\'enic or as
a super-Alfv\'enic tail flow. An example is our own heliosphere and
its heliotail.
}
{
We present an analytical method
of calculating stationary, incompressible, and field-aligned plasma flows in the
astrotail of a star. 
We present a recipe for constructing an astrosphere
with the help of only a few governing parameters, like
the inner Alfv\'en Mach number
and the outer Alfv\'en Mach number, the magnetic field strength
within and outside the stellar wind cavity, and the distribution of
singular points (neutral points) of the magnetic field within these flows.
}
{
Within the framework of a one-fluid approximation, it
is possible to obtain solutions of the governing MHD equations for stationary
flows from corresponding static MHD equilibria, by
using noncanonical mappings of the canonical variables. 
The canonical variables are the Euler potentials of the magnetic
field of magnetohydrostatic equilibria.
Thus we start from static equilibria determined by the 
distribution
of magnetic neutral points, and assume that the Alfv\'en Mach number for the
corresponding stationary equilibria is finite.
}
{
The topological structure, i.e. the distribution of magnetic
neutral points, determines the geometrical structure of the
interstellar gas - stellar wind interface.
Additional boundary conditions like 
the
outer magnetic field and the jump of the
magnetic field across the astropause
allow determination of the
noncanonical transformations.
This delivers the strength of the 
magnetic field at
every point in the astrotail/astrosheath region beyond the reverse shock.
}
{The mathematical technique for describing
such a scenario is applied to 
astrospheres
in general, but is also relevant for the heliosphere. It shows the 
restrictions of the outer and the inner magnetic field strength in comparison
with the corresponding Alfv\'en Mach numbers in the case of subalfv\'enic
flows.} 


\titlerunning{MHD flows in astrotails}
\authorrunning{D.H. Nickeler et al.}
\maketitle

\section{Introduction}

  \subsection{The scenario}

A wide range of literature is concerned with calculating stationary
MHD flows for stellar magnetospheres, jets, stellar winds, and laboratory
or general plasma configurations, see e.g. Chandrasekhar (\cite{Chandra}),
Tsinganos (\cite{Tsinganos}), Lovelace et al. (\cite{Love}), and
Goedbloed \& Lifschitz (\cite{Goedbloed}).
These authors use one flux function to represent two components of the
magnetic field and get a Grad-Shafranov type equation. This is a single
nonlinear, partial differential equation for this magnetic flux function,
and the method is restricted to 2D fields and flows. We apply
a method that is not restricted to one flux function, but works with two
flux functions of the magnetic fields.

\begin{figure}
\resizebox{\hsize}{!}{\includegraphics{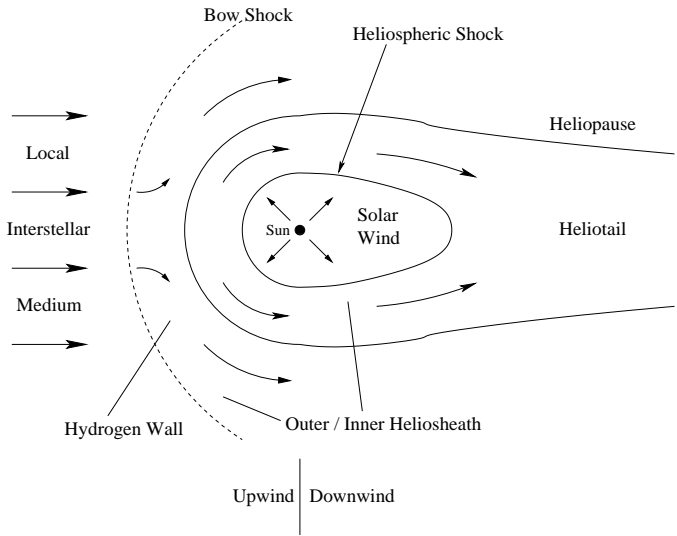}}
\caption{Sketch of the heliosphere shown as a special example of an
astrosphere.}
\label{scenario}
\end{figure}

Here we show that, under certain reasonable assumptions,
it is possible to use a powerful transformation method for systematic modelling
of the stellar wind region far away from the star itself. We apply this method
to the special scenario of a stellar wind--interstellar medium (ISM)
counterflow configuration (see Fig.\,\ref{scenario} for the special case
of the heliosphere).
 From this scenario, it is possible to estimate the pressure of the magnetic
field,
plasma pressure, and ram pressure, which are dynamically important 
for the ISM and
therefore of high interest to astronomy (see e.g. Frisch\,\cite{Frisch}).

Beyond the region of the reverse shock (`Heliospheric Shock'
in Fig.\,\ref{scenario}), the plasma of the stellar wind is decelerated. 
The magnetic field
increases at the shock, so that a sub- or a 
super-Alfv\'enic plasma
flow exists in the downstream direction.
A contact surface forms between the two different flows.
This separatrix is called the astropause.
Scherer et al.\,(\cite{Scherer}) and Fahr et al.\,(\cite{Fahr4})
showed that the bulk flow in the downwind 
direction can be assumed
to be incompressible for small Mach numbers.
This holds even more for field aligned flows, since the field lines act as
quasi-isothermals. As the decelerated stellar wind has to adapt to 
the conditions
of an outer
magnetized {\bf V}ery {\bf L}ocal {\bf I}nter{\bf S}tellar {\bf 
M}edium (VLISM),
a tangential discontinuity forms. This is called the astropause (AP, 
indicated by `Heliopause' in Fig.\,\ref{scenario}   ),
which stretches out in the downwind direction, so that the whole 
structure has a
tail-like shape
(see Fig.\,\ref{scenario}). Between the termination shock (TS, the heliospheric
shock in Fig.\,\ref{scenario})
and the AP an inner astrosheath region extends into 
the astrotail,
similar to the Earth's magnetotail. In the magnetotail, the MHD 
quantities depend mainly on the direction perpendicular to the tail axis (e.g.
Schindler\,\cite{Schindler}).
In view of the evident similarities, we here apply a similar description to
astrotails.
We show that there is a strong correlation between the flow, especially
its Alfv\'en Mach number inside and outside the astropause region, and the
current density in the vicinity of the astropause. This correlation 
also sets
restrictions on the relation between outer and inner values of the magnetic
field and the Alfv\'en Mach numbers.

  \subsection{Model considerations and existing models}

Astrospheric models can be divided into two kinds, analytical
and numerical, but there is a gap between these two. In most cases,
the analytical models are only hydrodynamical or
purely magnetic models (e.g. Parker\,\cite{Parker}).
Parker's article only treats purely
hydrodynamic models of the subsonic counterflow or of an
unmagnetized stellar wind that blows into a
magnetohydrostatic interstellar environment (Parker\,\cite{Parker}).
His models connect parameters, like the pressure of
the magnetic field and the thermal pressure
, with the shape of the model-astrospheres.
Another analytical model was the first super-Alfv\'enic MHD model
suggested for the SW--ISM interaction, which is based on the thin-layer
(hypersonic) approximation (Baranov \& Krasnobaev\,\cite{Krasno}).
In contrast to the model of Parker
or Baranov and Krasnobaev
the geometry of our model, presented here in this article,
does not explicitly depend on the Alfv\'en Mach number or the
usual Mach number of the flow. We propose a different point
of view in this article by emphasizing
the connection between topological aspects of the magnetic field
structure and the geometrical shape of the astropause.

Other analytical models have been
calculated for pressure equilibrium (Newtonian approximation by
Fahr et al.\,\cite{Fahr2}) or for those configurations
where the plasma cavity is a finite
ellipsoid, and the plasma has to leave the astrosphere by diffusion
(Neutsch \& Fahr\,\cite{Neutsch}).
Other authors use the set of MHD equations, but prefer solving
the ideal Ohm's law
and neglecting the Lorentz force in the Euler equation
(see the kinematical approach by Suess \& Nerney\,\cite{Suess};\,\cite{Suess2};
Nerney et al.\,\cite{Nerney}).
Up to now no analytical and exact solutions of the MHD
equations exist for this scenario.
This motivated us to consider models that do not depend on mathematical
approximations, but where (additional) physical approximations 
are taken into account. For example, Imai (\cite{Imai}) analysed
field-aligned flows and calculated approximative solutions.
However, we take into account that a flow that has passed a shock is likely
to develop a sub-sonic/sub-Alfv\'enic flow with a negligible compressibility
along the field- and streamlines.
The argument in Scherer et al.\,(\cite{Scherer})
and Fahr et al.\,(\cite{Fahr4}) is that the low
Mach number does not provide high compressibility rates.
Another argument for \lq incompressibility\rq~is that in a
tail that is symmetric with respect to the tail axis,
the far-away field- and streamlines are nearly one-dimensional.
This implies a one-dimensional dependence of the physical
values perpendicular to this axis, see Schindler\,(\cite{Schindler}),
so that the density is approximately constant on field lines.

Webb et al.\,(\cite{Webb}) analysed two-dimensional MHD flows and discussed
the properties of transsonic flows.
We focus on the relation between the Alfv\'en Mach number
$M_A$ and the electric current density,
which can be given by the jump of the two magnetic fields across
the boundary of the astropause of, in principle, 3D configurations.
We also focus on
topological and geometrical questions with respect to the boundary
between two magnetized flows.

Simulation results may also be reliable, although they
can deliver unphysical results, even when stable algorithms are used.
For example, Linde et al.\,(\cite{Linde})
discovered magnetic diffusion in their simulation domain, although they
used an ideal MHD code. Evidently, numerical magnetic reconnection is
taking place in some region of their domain of calculation.

Numerical reconnection cannot take place in our treatment since
we calculate exact
and analytical solutions of the ideal MHD equations with finite
width  of the astropause current sheet. We describe
smooth flows without shocks, i.e. without non-tangential
discontinuities, only. In our approach, we have
restricted the analysis to tangential discontinuities. The possibility of
Alfv\'enic discontinuities is discussed by Smith\,(\cite{Smith})
for the heliospheric
current sheet.
Thus, we close the gap between the older analytical, but simplified,
treatments of the
astrophysical counterflow problem and the sophisticated numerical models.
We find principles for these counterflow scenarios that should also hold
for simulations
having fewer physical simplifications.
{\sl Our aim is to calculate analytical and exact solutions of the stationary
nonlinear MHD equations
    for stellar wind tails, extending the approximative analytical 
models discussed
above.}

The paper is structured as follows.
In Sect.\,2, we present the equations to be solved for our scenario
and the special method used to solve them.
In Sect.\,3, we show how this method works in the two-dimensional 
case, which is a
simplification of our model, focussing on the method and the basic 
principles of
MHD counterflows.
In Sect.\,4, we show how to construct the flow and its stream lines, i.e. the
magnetic field
structure. This gives us the pattern of such MHD flows. We use the simplest
magnetohydrostatic
equilibria, viz. potential fields. The reason for it will be given in
Sect.\,4.
Section\,5 discusses the dependence on the symmetry and boundary
conditions for the asymptotical one-dimensional case, which is interesting for
tail-like structures of the magnetic field.
In addition, we present some two-dimensional tail models. Discussion and
conclusions are given in Sect.\,6.

\section{Stationary states in incompressible and ideal MHD}

  The set of equations that must be solved to get
incompressible ideal MHD flows consists of the
mass continuity equation (\ref{mce}),
the Euler or momentum equation with isotropic pressure $P$ (\ref{ee}), the
induction equation including the ideal Ohm's law (\ref{ie}), Amp\`ere's law
(\ref{al}), the initial condition for the magnetic
field (\ref{sc}), and the condition for incompressibility (\ref{ice}):
\begin{eqnarray}
  \vec\nabla\cdot\left( \rho\vec{\rm v} \right) &=& 0\, ,\label{mce}\\
   \rho\left( \vec{\rm v}\cdot\vec\nabla\right)\vec{\rm v} &=& \vec 
j\times\vec B -
\vec\nabla P\, \label{ee},\\
    \vec\nabla\times\left(\vec{\rm v}\times\vec B\right) &=&\vec 0\, 
,\label{ie}\\
     \vec\nabla\times\vec B &=& \mu_{0}\vec j\, ,\label{al}\\
      \vec\nabla\cdot\vec B &=& 0\, ,\label{sc}\\
       \vec\nabla\cdot\vec{\rm v} &=& 0\,\label{ice} .
        \end{eqnarray}
Due to the incompressibility, the mass continuity equation can be written as
$\vec {\rm v}\cdot\vec\nabla\rho=0$, so that the density is constant on
streamlines.
If we now introduce the auxilliary flow vector $\vec 
w:=\sqrt{\rho}\,\vec{\rm v}$
and
the Bernoulli pressure $\Pi:=P+\frac{1}{2}\vec w ^2$, we can write the above
equations as
  \begin{eqnarray}
   \vec\nabla\cdot\vec w &=& 0\, ,\label{divwfree}\\
    \vec\nabla \Pi  &=&\frac{1}{\mu_{0}}\left( \vec\nabla\times\vec
B\right)\times\vec B -
    \left( \vec\nabla\times\vec w\right)\times\vec w\, ,\label{equ}\\
     \vec\nabla\times\left(\frac{1}{\sqrt{\rho}}\,\vec w\times\vec B\right) &=&
\vec 0\, ,\label{indeq} \\
      \vec\nabla\cdot\vec B &=& 0\, \label{solenoid}.
       \end{eqnarray}
Hence, the momentum equation Eq.\,(\ref{equ}) is written such that the analogy
with magnetohydrostatic equilibria,
$\vec\nabla P = \mu_{0}^{-1}\left( \vec\nabla\times\vec B\right)\times\vec
B$,
is evident.

\subsection{Field-aligned flows}

The stationary equilibria should be constructed such 
that they tend to be stable in
order to use them as stationary background fields in very turbulent and
time-dependent stellar winds.
  In analytical works (e.g. Suess \& Nerney\,\cite{Suess} and 
references therein),
this
problem is often treated kinematically, which means that the Lorentz force is
ignored.
These authors find strong amplification of convected magnetic fields
in the so-called
upwind direction, which is the direction from which the
interstellar medium is flowing
towards the star. In this direction they identified a cone of 30 degrees (where
the star is sitting at
the top of the cone), in which their
kinematical approach is invalid. Such velocity fields with a strong 
perpendicular component
to the magnetic field have a saddle-point structure in linear stability
analyses (Hameiri\,\cite{Hameiri}) and are, therefore, likely to develop
ideal MHD instabilities. 

Hameiri found that a variational principle
does not lead to a stability criterion if velocity and magnetic field are not
aligned, because the used functional has only stationary points, but has no
minimum.
  Hameiri (\cite{Hameiri}) suggests that the lack of a minimum is due to the
presence of ballooning modes.
In fact assuming the incompressible limit, the equilibrium velocity
field has to be sub-Alfv\'enic to ensure the existence of a minimum.
Thus, to calculate magnetohydrodynamic configurations that should \lq
survive\rq~ long enough to represent a quasi-stationary state of
the stellar wind flow,
it is necessary to assume field aligned flows. This ensures that 
these configurations
can really exist in nature, i.e. that they are sufficiently long-lived
to be represented as stationary MHD flows.
This would not be the case with strong
perpendicular components of the flow with respect
to the magnetic field, since those would lead to quick ideal MHD instabilities.
Thus, models where magnetic and flow field are not (approximately)
aligned cannot
exist in nature without showing strong time dependency.
Since strong time-dependent MHD flows,
instabilities, and shocks do occur in the corotating interaction regions
of the solar system, the validity
of our models is restricted to those regions far outside the termination
shock. There, incompressible and field-aligned flows are good
approximations of the real outflows of
stars, at least when they are sufficiently
far away from the stellar surface\footnote{
This is also valid,
if we focus our view either on the classical (subsonic
unmagnetized flow, radially extrapolated to the origin) Parker flow
or on the Parker spiral magnetic field, which we use as a
geometrical pattern for calculating
flows and magnetic fields in the next sections.}.

 From this point of view,
to lower the risk of instability, it is expedient
to make the simplifying assumption of field aligned flow, i.e.
\begin{equation}
  \vec w = \pm\,\frac{M_{A}}{\sqrt{\mu_{0}}}\,\vec B\, ,\label{aligned}
   \end{equation}
  where $M_{A}$ is the Alfv\'en Mach number. This equation fulfills 
the induction
equation
  (\ref{indeq}) automatically.
  The sign on the righthand side of Eq.\,(\ref{aligned}) is to be understood in
the framework of the
  transformation method introduced in the next subsection.
  With this assumption, we can skip Eq.\,(\ref{indeq}), and
%
%
%
from Eqs.\,(\ref{aligned}),\,(\ref{divwfree}), and \,(\ref{solenoid})
it follows directly that
  \begin{eqnarray}
   \vec w\cdot\vec\nabla M_{A} = 0\, ,\qquad\textrm{and}\qquad\vec 
B\cdot\vec\nabla
M_{A} = 0 \, .
    \end{eqnarray}
   Therefore, the mass density $\rho$ and the Alfv\'en Mach number are 
constant on
field lines,
   but they can vary perpendicular to them.
   In conclusion, we have to solve the following system of equations:
   \begin{eqnarray}
    \vec B\cdot\vec\nabla M_{A} &=& 0\, ,\label{solenoid3}\\
     \vec\nabla \Pi  &=&\frac{1}{\mu_{0}}\left(1-M_{A}^2\right)\left(
     \vec\nabla\times\vec B\right)\times\vec B - \frac{|\vec B|^2}{2\mu_{0}}
     \,\vec\nabla\left( 1-M_{A}^2\right)\, ,
     \label{equ2}\\
      \vec\nabla\cdot\vec B &=& 0\, . \label{solenoid4}
       \end{eqnarray}
   This system determines the unknowns $\vec B$, $M_{A}$ and $\Pi$.
   In the next subsection we present a method of solving
   the general three-dimensional problem given by the system of
   Eqs.\,(\ref{solenoid3})--(\ref{solenoid4}) by means of a noncanonical
transformation.
Later on, we explicitly calculate two-dimensional
equilibria in order to highlight the main properties that are important
for understanding relaxed,
magnetized, stellar tail flows.
The reduction of this system to one magnetic flux function
(e.g. Tsinganos \cite{Tsinganos}; Goedbloed \&
Lifschitz\,\cite{Goedbloed}) will be done in Sect.\,\ref{gse}.

\subsection{Euler potential representation and noncanonical transformations}

In most cases in the literature (e.g. Chandrasekhar\,\cite{Chandra};
Tsinganos\,\cite{Tsinganos}; Lovelace et al.\,\cite{Love};
Goedbloed \& Lifshitz\,\cite{Goedbloed}), the problem of solving the
stationary MHD equations is reduced
to equations similar to the Grad-Shafranov equation (GSE, see e.g. 
Grad \& Rubin\,\cite{Grad}) by introducing two-dimensional flux functions
for the magnetic field.
Here, we give a short introduction to a different method that allows 
us to calculate also fields that could be three dimensional.

In 1984, Zwingmann showed the similarity between magnetohydrostatic (MHS)
equilibria and stationary
MHD equilibria with incompressible, field-aligned flows.
Later, this theory was improved by
Gebhardt \& Kiessling\,(\cite{Geki}), and subsequently used
by Petrie \& Neukirch\,(\cite{Petrie}) 
for modelling sunspot magnetic fields with plasma
flow. We briefly recapitulate the
transformation method used in the cited
papers in order to facilitate the analysis of our astrospheric model.

In general, non-ergodic magnetic fields can be represented
by using Euler potentials
(see e.g. Kruskal \& Kulsrud\,\cite{Kruskal} or D'haeseleer\,\cite{Dhaeseleer},
and references therein).
The magnetic fields of MHS equilibria can also be represented
by using Euler potentials, writing
\begin{equation}
  \vec B_{S}=\vec\nabla f\times\vec\nabla g ,
  \end{equation}
  where the Euler or Clebsch potentials $f$ and $g$ are scalar functions
  of $x$, $y$, $z$ in general.
Here, and in the following, the subscript $S$ will be used to indicate
magnetohydrostatic equilibrium quantities.
  The MHS equations can now be written as canonical Hamiltonian
  equations:
  \begin{eqnarray}
   \frac{\partial P_{S}}{\partial f} &=& \vec j_{S}\cdot\vec\nabla g ,\\
    \nonumber\\
      - \frac{\partial P_{S}}{\partial g} &=& \vec j_{S}\cdot\vec\nabla f ,
    \end{eqnarray}
  with the canonical variables $f$ and $g$ and the arc length $s$ 
along the current
$j_{S}$
  (see Schindler\,\cite{Schindler2}).
The MHS field $\vec B_{S}$ can now be mapped to a new field $\vec B$ by
performing the transformation,
\begin{eqnarray}
   f &=& f(\alpha,\beta) \quad\qquad\qquad\quad \alpha=\alpha(f,g)\nonumber\\
   &&\qquad\qquad\quad\Longleftrightarrow\nonumber\\
   g &=& g(\alpha,\beta) \quad\qquad\qquad\quad \beta =\beta(f,g)\,\,
\label{trafomag}
     \end{eqnarray}
where the derivatives as well as the inverse mappings and its derivatives,
are assumed to exist.
Then, there is a relationship between the old (static) field $\vec 
B_{S}$ and the
new field $\vec B$, which
  can be interpreted as a stationary field:
\begin{equation}
  \vec B_{S}=\vec\nabla f\times\vec\nabla
g=[f,g]_{\alpha,\beta}\vec\nabla\alpha\times\vec\nabla\beta\equiv
  [f,g]_{\alpha,\beta}\,\vec B\, ,\label{trafomagneu}
  \end{equation}
where the Poisson bracket $[f,g]_{\alpha,\beta}$ is the Jacobian of the
transformation Eq.\,(\ref{trafomag}).
If the Poisson bracket $[f,g]_{\alpha,\beta}= 1$,
it can be seen from Eq.\,(\ref{trafomagneu})
that only a change of the potentials but no real active
transformation has taken place,
so that the magnetic field has not been changed.
Therefore, $\alpha$ and $\beta$ would also be canonical
variables for the field $\vec B_{S}$, and the mapping from $\vec 
B_{S}$ to $\vec B$ would be a canonical
transformation, which does not produce new physics. However, if the Poisson
bracket
  $[f,g]_{\alpha,\beta}\neq 1$, then the magnetic field $\vec B$
  has to be interpreted in a different way, as it is not possible to 
identify it as
a magnetic field
  of a static equilibrium.

  The similarity between MHS, Eq.\,(\ref{mhs}) below, and MHD, 
Eq.\,(\ref{equ2}),
can be seen by inspecting
the original momentum balance equation of the MHS field, when we insert
  Eq.\,(\ref{trafomag}) into $\vec\nabla P_{S}=\vec\ j_{S}\times\vec 
B_{S}$, which
leads to
  \begin{eqnarray}
\vec\nabla P_{S}=&&\frac{1}{\mu_{0}}\left(\vec\nabla\times\left(\vec\nabla
f\times\vec\nabla
     g\right)\right)\times\left(\vec\nabla f\times\vec\nabla
   g\right)\, \nonumber\\
    =&&\frac{[f,g]_{\alpha,\beta}^2}{\mu_{0}}\,\vec\nabla\times\left(
    \vec\nabla\alpha\times\vec\nabla\beta\right)
    \times\left(\vec\nabla\alpha\times\vec\nabla\beta\right)
    \, \nonumber\\
       &&-
\frac{\left|\vec\nabla\alpha\times\vec\nabla\beta\right|^2}{2\mu_{0}}\vec\nabla
[f,g]_{\alpha,\beta}^2
        \label{mhs}\, .
         \end{eqnarray}
Consequently, the relation between the Poisson bracket and the Alfv\'en Mach
number is given by
  \begin{equation}
   0<\left(\left[f,g\right]_{\alpha,\beta}\right)^2:=1-M_{A}^2\label{sub}
     \end{equation}
for purely sub-Alfv\'enic flows, and by
  \begin{equation}
   0<\left(\left[f,g\right]_{\alpha,\beta}\right)^2:=M_{A}^2-1\label{super}
     \end{equation}
for purely super-Alfv\'enic flows.

Setting $\vec B = \vec 0$ in Eq.\,(\ref{equ}),
which describes a purely incompressible stationary hydrodynamical flow,
Gebhardt \& Kiessling (\cite{Geki}) noted the similarity between
Eq.\,(\ref{equ}) and Eq.\,(\ref{mhs}).
This can be seen if the auxiliary flow field $\vec w$ is also 
represented by Euler
potentials.
Then a mapping from a known solution of stationary incompressible 
hydrodynamics to
a stationary
incompressible, super-Alfv\'enic field-aligned flow is possible, if
$([f,g]_{\alpha,\beta})^2 > 1$ .
It is possible to map a known solution of the MHS 
equations by means of a
transformation with  $([f,g]_{\alpha,\beta})^2 < 1$ to an incompressible MHD
equilibrium with a field-aligned sub-Alfv\'enic flow.
Thus for every incompressible field-aligned sub-Alfv\'enic flow,
it is possible to find a mapping onto a MHS equilibrium.
Alternatively, it is also possible to take a different look at the problem
by focussing on the \lq current-generating\rq\, transformation of a given MHS
equilibrium, as we are interested in the astropause current sheet. This
works as follows:
from Eqs.\,(\ref{mhs}) and\,(\ref{super}) it is obvious that for a certain
transformation,
e.g. given by a sub-Alfv\'enic flow with the Jacobian squared
$([f,g]_{\alpha,\beta})^2 < 1$,
   \begin{eqnarray}
    \vec w_{sub} &=& \textrm{sign}\left[\left[f,g\right]_{\alpha,\beta}\right]\,
     \frac{\sqrt{1 -
\left(\left[f,g\right]_{\alpha,\beta}\right)^2}}{\sqrt{\mu_{0}}}\,\vec B\,
,\label{wsub}
     \end{eqnarray}
there exists a corresponding super-Alfv\'enic solution for the flow fields:
    \begin{eqnarray}
     \vec w_{super} &=& 
\textrm{sign}\left[\left[f,g\right]_{\alpha,\beta}\right]\,
     \frac{\sqrt{\left(\left[f,g\right]_{\alpha,\beta}\right)^2 +
1}}{\sqrt{\mu_{0}}}\,\vec B\, .\label{wsuper}
      \end{eqnarray}
  This has to obey the following restriction
  \begin{eqnarray}
     0 &<& [f,g]_{\alpha,\beta}^2=M_{A,super}^2-1 \equiv 1- M_{A,sub}^2 <
1\label{msub}\\
  \Rightarrow && M_{A,super}^2 \equiv  2 - M_{A,sub}^2  \, ,\label{msuper}
    \end{eqnarray}
to be satisfied at every point in space, while the magnetic field is the same
as in the sub-Alfv\'enic case:
$\vec B_{super}\equiv\vec B_{sub}\equiv \vec B$.

Therefore, it is not guaranteed
for all Poisson brackets, i.e. transformations,
that sub-Alfv\'enic solutions
exist, but for a given sub-Alfv\'enic solution $([f,g]_{\alpha,\beta})^2 < 1$
a corresponding super-Alfv\'enic MHD flow exists
with $1 < M_{A,super}^2 < 2 $. For
these flows, the magnetic field of the underlying MHS
equilibrium magnetic field will be amplified:
  \begin{equation}
   \left|\vec B \right|\equiv \left|\vec B_{super}\right| =
   \left|\frac{\vec B_{S}}{\sqrt{M_{A,super}^2 - 1}}\right| > \left|\vec
B_{S}\right|\, .\label{bsuperin}
    \end{equation}

  It is also necessary for the super-Alfv\'enic case that the thermal or plasma
pressure $\Pi_{super}$ is
  \,\lq inverted\rq\, to regain the similarity between Eqs.\,(\ref{equ2})
and\,(\ref{mhs}),
\begin{eqnarray}
  \vec\nabla P_{S}\equiv\vec\nabla\Pi_{sub}\, ,
  \end{eqnarray}
  changed to
  \begin{eqnarray}
  \vec\nabla P_{S}\equiv\vec\nabla\left( - \Pi_{super}\right)\, ,
  \end{eqnarray}
where $\Pi_{sub}$ is the sub- and $\Pi_{super}$ is the corresponding 
super-Alfv\'enic Bernoulli pressure.
Integration of these equations leads to
  \begin{eqnarray}
    P_{S} &=& \Pi_{sub} + \Pi_{0}\label{pisub}\, ,\\
   P_{S} &=& -\Pi_{super} + \Pi_{1}\label{pisuper}\, ,
     \end{eqnarray}
where $\Pi_{0}$ and $\Pi_{1}$ are integration constants. It follows from
Eqs.\,(\ref{pisub}) and
(\ref{pisuper}) that
\begin{eqnarray}
  P_{super} = \Pi_{1} - \Pi_{0} - \left(\frac{|\vec B|^2}{2\mu_{0}} +
P_{sub}\right)\, .\label{pesuper}
    \end{eqnarray}
All the above-mentioned relations and considerations are also valid 
for the case
$M_{A,super}^2 - 1 = ([f,g]_{\alpha,\beta})^2 > 1$. Only
Eqs.\,(\ref{msub}),\,(\ref{msuper})
,\,(\ref{pisub}), and (\ref{pesuper}) are not valid then, along with 
the inequality Eq.\,(\ref{bsuperin}).
However, what does remain valid is that the magnetic field of the
underlying MHS equilibrium can be amplified or weakened:
  \begin{equation}
   \vec B\equiv\vec B_{super} = \frac{\vec B_{S}}{\sqrt{M_{A,super}^2 - 1}} ,
\label{bsuper}
    \end{equation}
  with domains where $|\vec B|$ can be larger, and domains where 
$|\vec B|$ can be
smaller than $\vec B_{S}$.

We want to focus on purely sub-Alfv\'enic flows.
In this case the transformation equations from the static to the 
stationary fields
can be written as
  \begin{eqnarray}
   \vec B_{S} &=&\vec\nabla f\times\vec\nabla g\quad\longmapsto\quad \vec
B=\vec\nabla\alpha\times
   \vec\nabla\beta=\frac{\vec B_{S}}{\sqrt{1-M_{A}^2}}\, ,\\
    P_{S} &=& P_{S}(f,g)\,\,\quad\longmapsto\quad P=P_{S} - \frac{1}{2}\rho
   |\vec{\rm v}|^2\, ,\\
     \vec{\rm v} &=& 0\qquad\quad\quad\longmapsto\quad \vec{\rm v} =
\frac{M_{A}\vec B_{S}}{\sqrt{\mu_{0}\rho
     \left(1-M_{A}^2 \right)}}\, .
      \end{eqnarray}

We are interested in the fact that an astrosphere is terminated by a boundary
between two different magnetic
fields. For the magnetic field, this boundary is a tangential 
discontinuity or an
encounter of two magnetic fields with a large gradient across that boundary.
This boundary, the astropause, can therefore be regarded as a current layer,
so we need additional information. We can deduce that
the electric current density is also transformed by
  \begin{eqnarray}
    \mu_{0}\vec j &=&
\vec\nabla\times\left(\vec\nabla\alpha\times\vec\nabla\beta\right)\nonumber\\
    &=& \vec\nabla[\alpha,\beta]_{f,g}\times\left( \vec\nabla 
f\times\vec\nabla g
\right) +[\alpha,\beta]_{f,g}
    \vec\nabla\times\left( \vec
    \nabla f\times\vec\nabla g \right)\nonumber\\
    &=&  \vec\nabla[\alpha,\beta]_{f,g}\times\vec B_{S} +
\mu_{0}\,[\alpha,\beta]_{f,g}\vec j_{S}
    \nonumber\\
    \nonumber\\
     &\stackrel{M_{A}<1}{=}& \frac{\rm{sign}\left[\left[\alpha,\beta
\right]_{f,g}\right]\, M_{A}}
     {\sqrt{\left(1 - M_{A}^2\right)^3}}\left[
    \vec\nabla f \left( \vec\nabla M_{A}\cdot\vec\nabla g \right)
    - \vec\nabla g \left( \vec\nabla M_{A}\cdot\vec\nabla f \right)
\right]\nonumber\\
       && + \frac{\rm{sign}\left[\left[\alpha,\beta 
\right]_{f,g}\right]\,}{\sqrt{1
- M_{A}^2}}
       \,\vec\nabla\times
       \left( \vec\nabla f\times\vec\nabla g \right)\, ,
        \label{jtrans1}\end{eqnarray}
which implies that even in the case of a static equilibrium with 
vanishing current
density, i.e.
a potential field, one gets a stationary equilibrium with a non-vanishing
current,
  \begin{eqnarray}
     \vec j = \frac{\rm{sign}\left[\left[\alpha,\beta \right]_{f,g}\right]\,
M_{A}}{\mu_{0}\,\sqrt{\left(1 -
     M_{A}^2\right)^3}}
     \left[ \vec\nabla f \left( \vec\nabla M_{A}\cdot\vec\nabla g \right)
     - \vec\nabla g \left( \vec\nabla M_{A}\cdot\vec\nabla f \right) 
\right] \, .
\nonumber\\
        \label{jtrans2}
         \end{eqnarray}
In addition to the previous works by the mentioned authors (Gebhardt
\& Kiessling\,\cite{Geki}
; Petrie \& Neukirch\,\cite{Petrie}), we have found that the flow and
the current are strongly correlated by means of the Alfv\'en Mach number.

\section{Two-dimensional equilibria}\label{gse}

To get exact and analytical equilibria, we restrict our view to 
two-dimensional equilibria since only symmetric equilibria are known
in infinite domains, see e.g. Tsinganos\,(\cite{Tsinganos2}).
There are no analytical and exact 3D MHD equilibria 
known that are bounded and that extend throughout
the whole 3D space.
Therefore, we assume from now on that $f=A$, where $A$ is a
function of $x$ and $y$,
and that $g=z$, so that we get the GSE (see e.g. Grad \& Rubin\,\cite{Grad}):
\begin{equation}
    \Delta A = -\mu_{0}\,\frac{dP_{S}}{dA}= -\mu_{0}\, j_{zS} \, ,
     \end{equation}
and a relation between the Alfv\'en Mach number and the derivative of $\alpha$:
  \begin{equation}
    M_{A}^2=1-
\frac{1}{\alpha'(A)^2}\quad\Leftrightarrow\quad\left(\alpha'(A)\right)^2=\left(
\frac{d\alpha}{dA}
    \right)^2=\frac{1}{1-M_{A}^2}\, . \label{sub2}
   \end{equation}
The current density for the stationary equilibrium can then be expressed by
  \begin{eqnarray}
   \Delta\alpha=-\mu_{0} j_{z}=\vec\nabla\cdot\left(\vec\nabla\alpha
\right)=\alpha''(A)\,\left|\vec\nabla A
   \right|^2 + \alpha'(A)\Delta A \, .
     \end{eqnarray}
The calculation of the current can also be derived from 
Eqs.\,(\ref{jtrans1}) and (\ref{jtrans2})
where we have transformed the current
density for general 3D equilibria following 
Amp\`ere's law, setting $f=A$ and $g=z$.
This is reasonable, if a domain in the vicinity of the equatorial plane of the
star is to be represented by the calculated equilibria.

\section{The pattern of the flow field and the magnetic field: 
potential fields as
magnetohydrostatic
  equilibra}\label{pattern}
   
We show in this section that the distribution of (virtual or 
real) magnetic neutral
points, which are also stagnation points if we assume
a finite Alfv\'en Mach number,
determines the global topological and geometrical structure of the astrosphere.
In the linear case, there is a relation between the magnetic 
multipole moments and
the neutral points that will be given later on. But also for a more complicated
class of nonlinear solutions of the MHS equations, it is possible to 
find similar
relations. Those will be discussed in a future paper.

There are several reasons for using potential fields as origins for 
our mappings.
They are much simpler to handle than nonlinear MHS fields in the 
framework of our solution
technique, where the nonlinearity of the MHD equations is handled by a
nonlinear
mapping technique. Hence, one only has to solve a linear partial 
differential equation,
whereas the nonlinearity is hidden in a nonlinear algebraic equation.
In addition, potential fields have a highly stable character as they 
have no free
magnetic energy, although it is not known if stability is conserved
after
the mapping onto a stationary equilibrium (see Petrie \& 
Neukirch\,\cite{Petrie}).
Also, the connection between the neutral point distribution of the magnetic
field and the global structure of the magnetic and the velocity fields 
can clearly be seen.

Another reason is that there should be at least one saddle point 
(the so-called X-point)
in this counterflow configuration. Amongst ($N+1$) null points,
at least one point must exist in the vicinity of the nose of the astropause
(at the stagnation point) for which the eigenvalues of the
Jacobian of the linearized magnetic field
\footnote{The global coordinates $(X,Y)$ are replaced here by
$(x,y)=(X-X_S,Y-Y_S)$, with $(X_S,Y_S)$ being the coordinates of the
null point.},
\begin{equation}
\left(\begin{array}{cc}
\displaystyle\frac{\partial B_{x}}{\partial x}
&\displaystyle\frac{\partial B_{x}}{\partial y}\\
\\
\displaystyle\frac{\partial B_ {y}}{\partial x}
&\displaystyle\frac{\partial B_{y}}{\partial y}\\
\end{array}\right)
\displaystyle\left(\begin{array}{c}
  x\\
  y
\end{array}
\right)
=\displaystyle\left(\begin{array}{c}
B_{1x}\\
B_{1y}
\end{array}
\right)\, ,
\end{equation}
take the form
\begin{equation}
  \lambda=\pm \, \lambda_{1} ~\, ,\, \lambda_{1}\in\, {\textrm I 
\!\textrm R} \, .
  \end{equation}
  With the Grad Shafranov equation
\begin{equation}
  \Delta A=-\mu_{0}\,\displaystyle\frac{dP}{dA}=J(A)\quad\Rightarrow\quad
\frac{\partial^2 A}{\partial y^2}=J(A)-\frac{\partial^2 A}{\partial x^2}~\, ,
  \end{equation}
where $J(A)$ is the current function, we can write the
linearized magnetic field as

\begin{equation}
\left(\begin{array}{cc}
\displaystyle\frac{\partial\,^2 A}{\partial x\,\partial y}
&\displaystyle J(A)-\frac{\partial^2 A}{\partial x^2}\\
\\
\displaystyle -\frac{\partial\, ^2 A}{\partial x^2}
&\displaystyle -\frac{\partial\, ^2 A }{\partial x\,\partial y }\\
\end{array}\right)
\displaystyle\left(\begin{array}{c}
  x\\
  y
\end{array}
\right)
=\displaystyle\left(\begin{array}{c}
B_{1x}\\
B_{1y}
\end{array}
\right) \, .
\end{equation}
The properties of the eigenvalues of the Jacobian of the magnetic field
determine whether the neutral (or stagnation) point is a point with which
one can define a separatrix curve (or surface, see Arnold\,\cite{Arnold};
Reitmann\,\cite{Reitmann}).
The separatrix is a border surface between two different regions of a 
flow, or of a magnetic field that
separates a vector field in areas of different topological connections. To get
information about
the topological structure of the magnetic field, we calculate
\begin{equation}
\textrm{Det}
\left(\begin{array}{cc}
\displaystyle\frac{\partial\,^2 A}{\partial x\,\partial y}-\lambda
&\displaystyle J(A)-\frac{\partial^2 A}{\partial x^2}\\
\\
\displaystyle -\frac{\partial\, ^2 A}{\partial x^2}
&\displaystyle -\frac{\partial\, ^2 A }{\partial x\,\partial y }-\lambda\\
                          \end{array}\right)
=0
\end{equation}
\vspace{0.5cm}
\begin{eqnarray}
\Rightarrow\quad \lambda^2
-\left(\displaystyle\frac{\partial\,^2 A}{\partial x\,\partial y}\right)^2
-\left(\displaystyle\frac{\partial\, ^2 A}{\partial x^2}\right)^2
+J(A)\, \displaystyle\frac{\partial\, ^2 A}{\partial x^2}=0 \, .
\end{eqnarray}
%
Therefore, we obtain the following eigenvalues:
\begin{eqnarray}
\lambda &=&\pm\left[
\left(\displaystyle\frac{\partial\,^2 A}{\partial x\,\partial y}\right)^2
+\left(\displaystyle\frac{\partial\, ^2 A}{\partial x^2}\right)^2
-J(A)\, \displaystyle\frac{\partial\, ^2 A}{\partial x^2}
\right]^{\frac{1}{2}}\label{eigen1}\\
&=& \pm\left[
\left(\displaystyle\frac{\partial\,^2 A}{\partial x\,\partial y}\right)^2
+2\,\left(\displaystyle\frac{\partial\, ^2 A}{\partial x^2}\right)^2
+\displaystyle\frac{\partial\, ^2 A}{\partial y^2}\, 
\displaystyle\frac{\partial\,
^2 A}
{\partial x^2}
\right]^{\frac{1}{2}} \, .
\end{eqnarray}
For two conjugate complex solutions in 2D, we obtain
$\lambda=\pm\lambda_{1}\in\,$ {\small\sf l}
$\!\!\!\!\!\begin{mathbf}C\end{mathbf}$,
i.e. only two purely imaginary values exist.
If only two real eigenvalues exist, there is a saddle point (also called
X-point), which is necessary for the existence of a separatrix and which
guarantees that there is a boundary surface between two distinct areas of the
flow;
i.e. an astropause exists. If $A$ is a potential field,
  i.e. $\Delta A=J(A)=0$, then the null point is a saddle point, as can be
seen from Eq.\,(\ref{eigen1}).
If the eigenvalues at the null point are purely
  imaginary, then, depending on the absolute value, a 
{\sl centre} (a
so-called {\it
O-point}) exists
  with topological circles as fieldlines, or a so-called
{\it focus} with spiral fieldlines. The last case is not found for
solenoidal vector fields, since the trace of the
Jacobian matrix vanishes for them.
%
%

\subsection{Potential fields}

  We now have to calculate the flow pattern, i.e. the magnetic field 
pattern. This can be done by calculating
  the most simple and stable magnetic fields, namely potential fields. 
We construct solutions by using a 2D multipole representation
in the form of a Laurent series, which enables us to find
static equilibria.
For this general kind of a conformal mapping, we exclude the region around the
singularity
  $(x,y)=(0,0)$ with $\varrho=\sqrt{x^2+y^2}<R_{ts}$ within the 
termination shock,
representing the
  inner part of the astrosphere. We define $u:=x+iy$.
  ${\cal A}$ and ${\cal B}$ are the complex
  magnetic flux function and the complex magnetic field.
  ${\cal A}(u)$ and ${\cal B}(u)$ are holomorphic functions (at least nearly
everywhere), with a
real part $\Re({\cal A})=\phi_{m}$ and imaginary part
  $\Im({\cal A})=A$, and $\Re({\cal B})=B_{x}$
  and $\Im({\cal B})=-B_{y}$ accordingly.
  $A$ is the magnetic flux function
and $\phi_{m}$ is the scalar
  magnetic potential. Therefore, we can write
\begin{equation}
{\cal B}=\frac{d{\cal A}}{du}=\frac{\partial\phi_{m}}{\partial x}
+i\frac{\partial A}{\partial x}=B_{x}-iB_{y}\, .
    \end{equation}
We use the following {\em Ansatz} for the  magnetic field
\begin{equation}
  {\cal B}=B_{S\infty}+\sum_{\mu=1}^{\infty}\, c_{\mu} u^{-\mu}\, ,
   \end{equation}
to satisfy the asymptotical boundary conditions
\begin{equation}
\lim_{|u|\rightarrow\infty}{\cal B}=B_{S\infty}\, ,
\end{equation}
so
\begin{equation}
{\cal A}= B_{S\infty}u+C_{0}\ln{u}+\sum_{\nu=1}^{\infty}\, C_{\nu} u^{-
\nu}\label{laurent1}
    \end{equation}
is valid.

There is a similarity between the logarithmic
term of the hydrodynamical problem of a
circular flow and the radial or azimuthal part of a magnetic 
potential field, so
that we can write
\begin{equation}
  {\cal A}=B_{S\infty}u+\frac{\Gamma}{2\pi i}\,\ln{u}
+\sum_{\nu=1}^{\infty} C_{\nu}\, u^{-\nu}\, ,
  \end{equation}
or especially
\begin{eqnarray}
  {\cal A} &=& B_{S\infty}u+\frac{\Gamma_{0}\left( 
\cos{\beta_{0}}+i\sin{\beta_{0}}
\right)}
{2\pi i}\ln{u}
-|C_{1}|\frac{\left( \cos{\beta_{1}}+i\sin{\beta_{1}} \right)}{u}\nonumber\\
\nonumber\\
&& -|C_{2}|\frac{\left( \cos{\beta_{2}}+i\sin{\beta_{2}}
\right)}{u^2}+\textrm{terms of higher order}\, .
\label{speziell}
   \end{eqnarray}
%
The first term in the expansion is the homogenous part due to the asymptotical
boundary condition, which survives the noncanonical transformation. 
The second  term is the
circulation or monopole part due to a line current. If
$\Gamma=\Gamma_{0}\left( \cos{\beta_{0}}+i\sin{\beta_{0}}\right)
$ is real, then one has a typical counterflow configuration for the
hydrodynamcal circulation of a flow around a circular cylinder.
Here
$\Gamma_{0}=\mu_{0}I_{0}$, where $I_{0}$ is the line current.
The third term is a line-dipole part. In the case of pure hydrodynamics,
this part represents the radius $R^2\equiv |C_{1}|$ of a flow around a
circular cylinder, if $\sin{\beta_{1}}=0$.
The fourth term is the quadrupole part. Furthermore,
$|C_{1}|=2I_{2}a$ is the dipole moment, $I_{2}$ the current,
$a$ the half distance of the antiparallel line currents, and the product
$I_{2}a=$\,const,
while $I_{2}\rightarrow\infty$ and $a\rightarrow 0$.

Moving stars, together with their winds, can be regarded as obstacles 
in the stream of the ISM. Such counterflow
configurations lead to the formation of separating surfaces. On these surfaces,
stagnation points must exist where the flow velocity vanishes.
Stream lines passing through these stagnation points
are called separatrices because they separate stream lines of
different topological connection. They represent the borderlines between two
different flows.
To calculate the stagnation points\footnote{
For a regular configuration, in the sense of finite Mach number and density
distribution,
these nulls or magnetic neutral points are also stagnation points.}, one has to
solve the equation ${\cal B}=0$.
In the topological theory of fluids, this field structure is called an
X-point structure\footnote{Such points are
saddle points of the magnetic flux function.}.
However, application to the Parker field would imply the existence of an
additional singularity beyond the X-point, located at the origin of the
magnetized wind
plasma, i.e. at the location of the central star itself.
To generate a field similar to the hydrodynamic Parker flow of the heliospheric
flow field (Parker\,\cite{Parker}), it is necessary (i) that
$x_{SP}\equiv x_{N} (<0)$
is the magnetic neutral point (where the magnetic field vanishes and in our
treatment also the velocity field), (i) that for $\varrho\equiv\sqrt
{(x^2+y^2)}\rightarrow\, 0$ a radial field structure exists, and (iii)
that the flow field converges
asymptotically to a homogenous field for $\varrho\rightarrow\infty$.
The Parker flow field is only useful and valid far away from
the origin of the magnetized wind plasma of the central star. One 
reason is, of course, that the field strength towards the origin is 
diverging.
It turns out that
selection of a circulation with a nonvanishing imaginary part
is necessary. This enables the stellar wind to escape from the region of the
reverse shock, and it creates a radial field structure towards the origin.

Note: In our case, the Alfv\'en or the usual Mach number
does not determine the geometrical shape of the astropause, in contrast
to the discussion in Parker\,(\cite{Parker}). Hence, our
method cannot be compared directly with Parker's calculations.
The mapping technique will allow sub- and superalfv\'enic flows
to exist, although the streamline geometry does not change.
Thus, it is also possible to have open field
lines on both sides (upwind and downwind), like in the purely
magnetic model of Parker.

\subsection{Neutral points and multipole moments}
The complex magnetic field ${\cal B}$ can be calculated from
${\cal A}=\phi_{m}+iA$, with $\phi_{m}$ as the magnetic potential
\begin{equation}
{\cal B}=\frac{d{\cal A}}{du}\, ,
   \end{equation}
where the {\em Ansatz} of the series (\ref{speziell}) yields
\begin{equation}
{\cal B}=B_{S\infty}+\frac{\Gamma}{2\pi i}\,\frac{1}{u}-\frac{C_{1}}{u^2}-
\frac{2C_{2}}{u^3} \, .
\end{equation}

A direct analytical method of calculating the null points should be 
applied here in
the case
of a multipole representation with two non vanishing
multipoles.
We restrict our analysis to the first two multipoles to get the
magnetic null:
  \begin{equation}
  {\cal B}=B_{S\infty}+\frac{\Gamma}{2\pi 
i}\,\frac{1}{u}-\frac{C_{1}}{u^2}= 0 \, .
   \end{equation}
This gives a quadratic equation,
\begin{equation}
  u^2-\frac{i\,\Gamma}{2\pi\, B_{S\infty}}\, u-\frac{C_{1}}{B_{S\infty}}=0\, ,
  \end{equation}
having the solutions
\begin{eqnarray}
u &=&\frac{i\Gamma}{4\pi\, B_{S\infty}}\pm\sqrt{
\left(\frac{i\Gamma}{4\pi\,
B_{S\infty}}\right)^2+\frac{C_{1}}{B_{S\infty}}}\nonumber\\
\nonumber\\
&=& \frac{i\left(\Gamma_{r}+i\Gamma_{i}\right)}{4\pi\, B_{S\infty}}\pm
\sqrt{ \left(\frac{C_{1r}}{B_{S\infty}}
+\frac{\Gamma_{i}^2-\Gamma_{r}^2}{16\pi^2\,B_{S\infty}^2}
\right) +i\left(\frac{C_{1i}}{B_{S\infty}}-\frac{\Gamma_{r}\Gamma_{i}}{8\pi^2\,
B_{S\infty}^2}    \right)  }
\nonumber\\
\nonumber\\
&=&\frac{-\Gamma_{i}+i\Gamma_{r}}{4\pi\, B_{S\infty}}\pm
\sqrt{\frac{\sqrt{{\cal R}^2+{\cal I}^2}+{\cal R}}{2}}\pm \frac{{i\cal
I}}{\sqrt{2\sqrt{{\cal R}^2+{\cal I}^2}+2{\cal R}}   }\, ,\nonumber\\
\nonumber\\
   \end{eqnarray}
with
\begin{equation}
{\cal R} =
\frac{C_{1r}}{B_{S\infty}}
+\frac{\Gamma_{i}^2-\Gamma_{r}^2}{16\pi^2\,B_{S\infty}^2}, \qquad
{\cal I} =
\frac{C_{1i}}{B_{S\infty}}-\frac{\Gamma_{r}\Gamma_{i}}{8\pi^2\, 
B_{S\infty}^2}\, ,
   \end{equation}
where the indices $r$ and $i$ indicate the
real and imaginary parts of the coefficients.
Therefore,
\begin{eqnarray}
  x_{S1} &=&
-\frac{\Gamma_{i}}{4\pi\, B_{S\infty}}\pm\frac{1}{\sqrt{2}}\left(
\sqrt{\sqrt{{\cal R}^2+{\cal I}^2}+{\cal R}}\right)\, ,\nonumber\\
\nonumber\\
  y_{S1} &=&\frac{\Gamma_{r}}{4\pi\, B_{S\infty}}\pm
\frac{{\cal I}}{\sqrt{2\sqrt{{\cal R}^2+{\cal I}^2}+2{\cal R}}   }.
  \end{eqnarray}
The singular point is situated where field lines meet; here is the
beginning, i.e. the ending of several field lines.
The stagnation point also marks {\it one} contour, e.g. a certain value of $A$.
From that, we can calculate the equation of the astropause and the asymptotical
equation of the astropause, which delivers the diameter of the 
astrotail at infinity.
The second stagnation point, which is calculated for the case of the 
symmetric linedipole
with
$y_{S2}=0$,  $x_{S2}>0>x_{S1}$, and $x_{S2}>|x_{S1}|$,
can be considered as the radius of the inner astrosphere, i.e. as the
position of the termination shock in the direction of the astrotail.
We perform a systematic calculation of the magnetic neutral points, making
the general {\em Ansatz} of a Laurent series.
If the Alfv\'en Mach number is finite, the magnetic neutral points 
are identical
with the stagnation points. We analyse how the position of the stagnation
points of the interstellar counterflow influences
the field structure, especially the geometry of the astropause. We use the
logarithmic part and the homogeneous asymptotic boundary condition, together
with the assumption of a finite number of neutral points:
\begin{equation}
{\cal A}= B_{S\infty}u+C_{0}\ln{u}+\sum_{\nu=1}^{N}\, C_{\nu} u^{-\nu} \,
.\label{multipol}
   \end{equation}
Here $\nu=1$ is the dipole, $\nu=2$ the quadrupole, $\nu=3$ the octopole,
etc.
The magnetic null or neutral points $u_{k}$ are given by
\begin{equation}
  {\cal B}(u_{k})=\frac{d{\cal A}}{du}\,\biggr|_{u=u_{k}}=0 \, .
   \end{equation}
Thus, we have to find the null points of the polynomial
\begin{equation}
  u^{N+1}+\frac{C_{0}}{B_{S\infty}}u^{N}-\sum_{\nu=1}^{N}\,
\frac{\nu C_{\nu}}{B_{S\infty}} u^{N-\nu}=\prod_{k=1}^{N+1}(u-u_{k})=0\, .
   \end{equation}
With the help of Vieta's theorem of roots, we get
\begin{eqnarray}
C_{0} &=& -B_{S\infty}\sum_{k=1}^{N+1} u_{k}\, ,\label{vieta1}\\
C_{\nu} &=& (-1)^{\nu}\,\frac{B_{S\infty}}{\nu}\,
\displaystyle\sum_{\bigcup{\cal 
C}_{\nu+1}^{N+1}}\,\left(\prod_{u_{k}\in{\cal C}_{
\nu+1
}^{N+1}} u_{k}\right)\, , ~ 1\leq\nu\leq N \, .\label{vieta2}
\end{eqnarray}
The symbol ${\cal C}_{\nu+1}^{N+1}$ denotes combinations of the $u_{k}$. These
are the subset $\nu+1$ elements of the $N+1$ elements
(of the magnetic nulls). {\sl The distribution of the magnetic neutral points
determines
the global geometry and the topology of an astrosphere.}

\subsection{The equation of the astropause}

Regarding a symmetric and, with respect to the direction of 
interstellar medium,
closed astrosphere, we only take
neutral points on the $x$ axis into account.
The smallest, negative, $x$ value gives
\begin{equation}
A(u_{1})=A_{sep1}=A(x_{1},0)\quad\Rightarrow\quad A_{sep1}=0\, ,
\end{equation}
and, therefore,
\begin{equation}
  A(x,y)=A_{sep1u}=0
  \end{equation}
for the separatrix with $x<0$. However,
including the monopole term, we obtain
for the point where the separatrix (astropause) passes through the $y$ axis:
\begin{equation}
\lim\limits_{x\rightarrow -0} A(x,y)=A(x=-0,y=y_{D})=A_{sep1}\label{hp1}\, .
\end{equation}
where $y_{D}$ is the location
of the astropause, which is lying on the $y$ axis and, therefore,
is elongated parallel to the inner astrosphere; i.e. it
is positioned at the same $x$ coordinate of the star $(x_{star},
y_{star})=(0,0)$. Now, $y_{D}$ can be determined from Eq.\,(\ref{hp1}).
For $x>0$, we may calculate the curve of the astropause in
the $x{-}y$ plane as an implicit function of
  \begin{eqnarray}
   A_{H}(x,y) &=& A(x=-0,y=y_{D})+\pi\, C_{0}\nonumber\\
   &=& A_{sep1}+\pi\, C_{0}=A_{sep2}
    = \pi\, C_{0}\, ,
     \end{eqnarray}
taking the jump of the arcus tangens function into consideration.
With three neutral points, the magnetic flux function can be written as
\begin{equation}
  \Im({\cal A})=A=B_{S\infty}y+C_{0}\arctan{\left(\frac{y}{x} \right)}-C_{1}\,
\frac{y}{x^2+y^2}-C_{2}\,\frac{2xy}{\left(x^2+y^2\right)^2}\, .
   \end{equation}
\begin{figure}
\resizebox{\hsize}{!}{\includegraphics{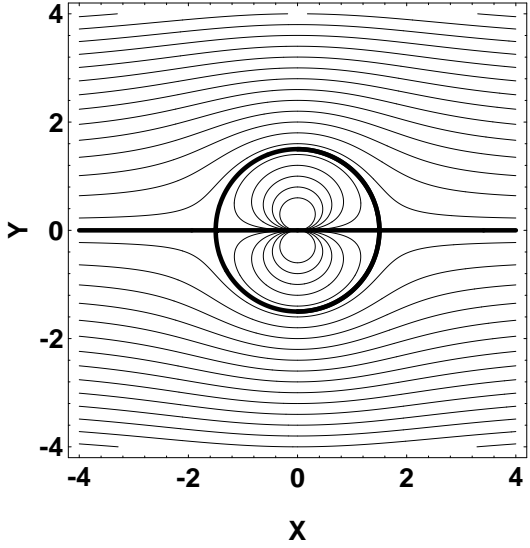}}
\caption{Field lines of the tail model as contour lines of the magnetic
           flux function and branches of the separatrix. $u_{1}$ is fixed at
           $-1.5$, where the scale is in
           units of 100\,AU. For $u_{2}=1.5$ one can clearly see
           the similarity of this magnetic field lines with that of a 
flow around
a cylinder.}
\label{hel1}
\end{figure}
\begin{figure}
\resizebox{\hsize}{!}{\includegraphics{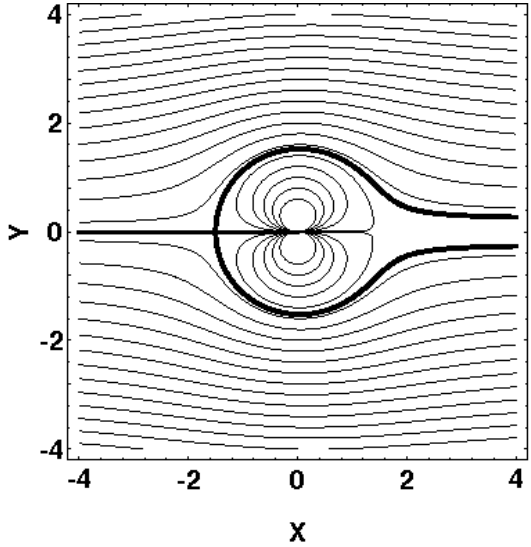}}
\caption{As in Fig.\,\ref{hel1}, but for $u_{2}=0.95\times 1.5=1.425$.}
\label{hel1b}
\end{figure}
For $x<0$ and
$y>0$ (with opposite sign for the limit of the
$\arctan$ for $y<0$, i.e. for the down part of the
astropause),
\begin{eqnarray}
  \lim\limits_{x\rightarrow\, -0}A(x,y)=A(0,y_{D})=B_{S\infty}y_{D}
-\frac{\pi}{2}\, C_{0}-\frac{C_{1}}{y_{D}}=0=A_{sep1}\, ,
  \end{eqnarray}
we get for the point where the separatrix intersects the $y$ axis
\begin{equation}
  y_{D}=\pm\left(\frac{\pi\, C_{0}}{4\, B_{S\infty}}+\sqrt{\left(\frac{\pi\,
C_{0}}{4 B_{S\infty}} \right)^2
  +\frac{C_{1}}{B_{S\infty}}}\,\right)\, .
    \end{equation}
\begin{figure}
\resizebox{\hsize}{!}{\includegraphics{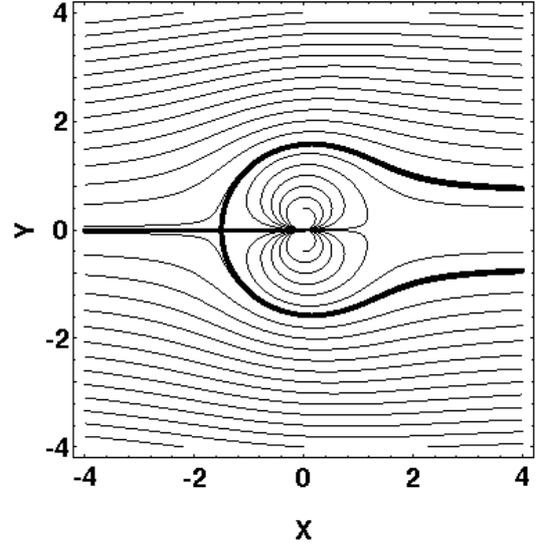}}
\caption{As in Fig.\,\ref{hel1}, but for $u_{2}=0.85\times 1.5=1.275$. 
That some field lines do not appear closed is only a plotting artifact.}
\label{hel2}
\end{figure}
\begin{figure}
\resizebox{\hsize}{!}{\includegraphics{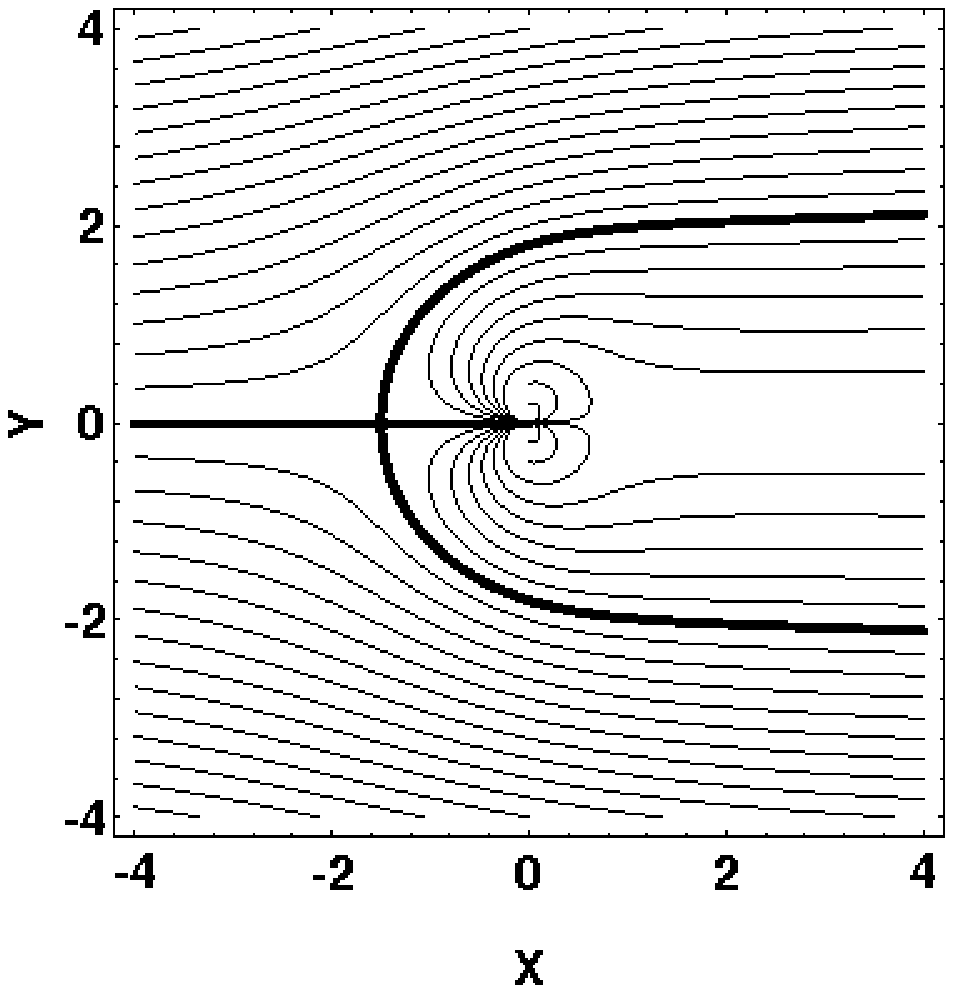}}
\caption{As in Fig.\,\ref{hel1}, but for $u_{2}=0.5\times 1.5=0.75$.
That some field lines do not appear closed is only a plotting artifact.}
\label{hel3}
\end{figure}
\begin{figure}
\resizebox{\hsize}{!}{\includegraphics{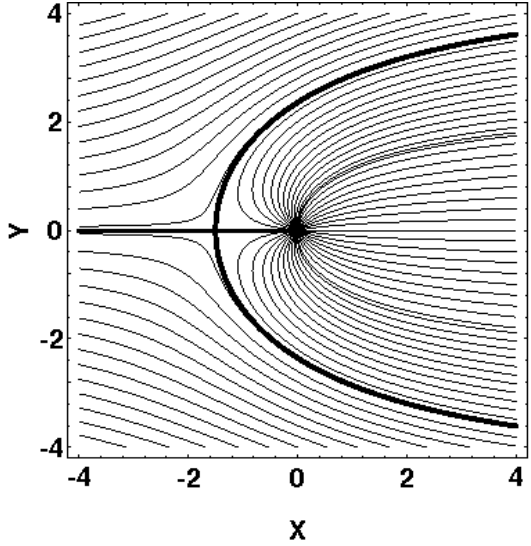}}
\caption{As in Fig.\,\ref{hel1}, but for $u_{2}=0$. In this special case 
the tail is similar to the Parker model (\cite{Parker}).
$u_{2}=0$ is not a null point, because this is the position of the
singularity. $u_{2}=0$ is only due to the fact that
in the sum of Eq.\,(\ref{multipol}) only a monopole moment appears;
see Eq.\,(\ref{vieta2}).}
\label{hel4}
\end{figure}

\begin{figure}
\resizebox{\hsize}{!}{\includegraphics{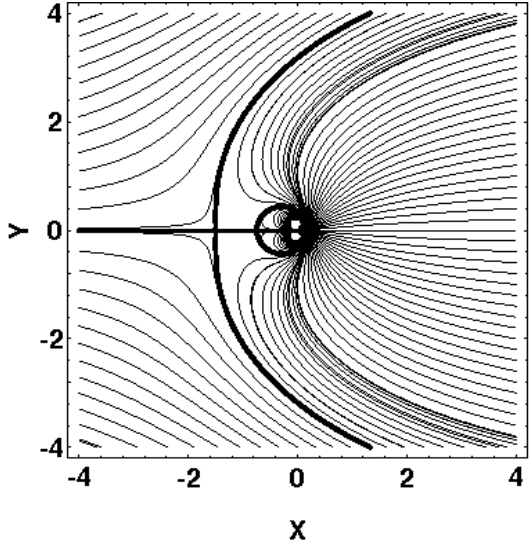}}
\caption{As in Fig.\,\ref{hel1}, but for $u_{2}=-0.5\times 1.5=-0.75$. 
Here, one gets a region with closed field lines in the
downstream direction.}
\label{hel5}
\end{figure}
\begin{figure}
\resizebox{\hsize}{!}{\includegraphics{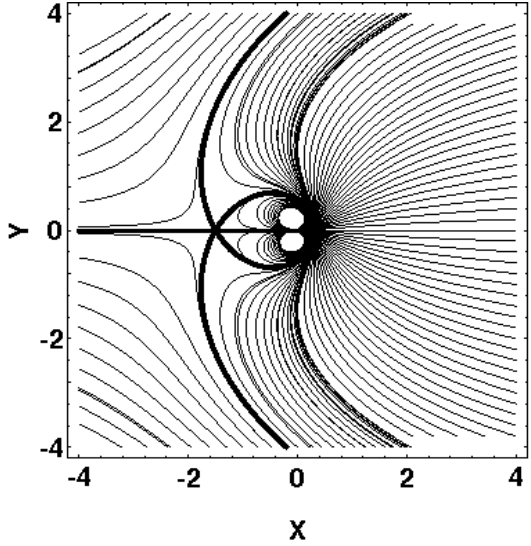}}
\caption{As in Fig.\,\ref{hel1}, but for $u_{2}=-1.5$;
           here we have a neutral point of second order.}
\label{hel6}
\end{figure}
\begin{figure}
\resizebox{\hsize}{!}{\includegraphics{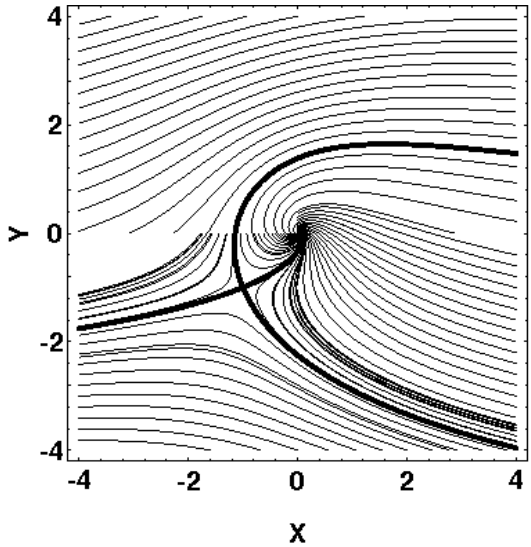}}
\caption{Field lines of the tail model as contour lines of the magnetic
   flux function and separatrices for $u_{1}=-1.0-i=(x,y)=(-1,-1)$;
   this is a potential field with a real valued circulation
   that shows the existence of a spiral-shaped field structure in the
   inner heliosphere, i.e. an azimuthal component;
   the thus imitated Parker spiral should extend into the outer heliosphere.
Due to a plotting artifact, not all calculated field lines are complete.}
\label{hel7}
\end{figure}

In the case of general asymmetric configurations, it is possible to calculate
the intersection point with the $y$ axis from
  \begin{eqnarray}
   \lim\limits_{x\rightarrow\, -0} A(x,y_D) &=& A(x_1,y_1)=A_{sep1}\nonumber\\
   &=& B_{S\infty}y_D-\frac{\pi}{2}C_{0r}+C_{0i}\,
   \ln{|y_D|}+\sum\limits_{\nu=1}^{N} C_{\nu}\, y_D^{-\nu}\, .
    \end{eqnarray}
With
\begin{equation}
  \lim\limits_{x\rightarrow\, +0}A(x,y_{D})=B_{S\infty}\,
y_{D}+C_{0}\,\frac{\pi}{2}
-\frac{C_{1}}{y_{D}}=A_{sep1}+\pi\, C_{0}=\pi\, C_{0}\, ,
   \end{equation}
we can calculate the other branch of the separatrix:
\begin{eqnarray}
A_H(x,y) &=& B_{S\infty}y_D+\frac{\pi}{2}C_{0r}+C_{0i}\,
\ln{|y_D|}+\sum\limits_{\nu=1}^{N} C_{\nu}\, y_D^{-\nu}\\
&\stackrel{here}{=}& B_{S\infty}y+C_{0}\arctan{\left(\frac{y}{x} 
\right)}-C_{1}\,
\frac{y}{x^2+y^2}-C_{2}\,\frac{2xy}{\left(x^2+y^2\right)^2}\nonumber\\
&=&\pi\, C_{0}\, ,
   \end{eqnarray}
which is the part of the astropause that is opened in the tail direction.
The outer separatrix is nothing else than the astropause.
The asymptotical equation of the astropause is given by
\begin{equation}
A_{H\infty}(x,y):=\lim\limits_{x\rightarrow\infty} A(x,y_{H})
=B_{S\infty}\, y_{H}=\pi\, C_{0}\,\,\Rightarrow
  \,\, y_{H}=\frac{\pi\, C_{0}}{B_{S\infty}}\, .
   \end{equation}
In the series of Figs.\,\ref{hel1} to \ref{hel7} we show how the 
existence of two null points
influences the shape of the astropause as an astrospheric interface. The 
scale is in
units of 100\,AU.
The first neutral point is chosen at a location that results in an astropause
configuration applicable to the heliosphere.

If we fix the neutral point, $x_{1}$, in
front of the astrosphere, we see that for $x_{2}=-x_{1}=R$ the
fieldline geometry looks like that of a hydrodynamical counterflow of a
cylinder with radius $R$ (Fig.\,\ref{hel1}).
By displacing the second null point towards the origin,
i.e. for $0<x_2<-x_1=R$, the separatrix breaks up, and a
tail-like channel is formed causing the drop-like shape of the astropause.
This tail opens up for a null point approaching the origin
(Figs.\,\ref{hel1b} - \ref{hel3}).
For $x_2 = 0$, the astrosphere has a typical Parker shape (Fig.\,\ref{hel4}).
Further displacement to negative values of $x_2$ results in the formation of an
anti-tailward bubble within the actual astropause (Fig.\,\ref{hel5}), leading
to a notch in the astropause nose. This bubble grows until
it touches the first neutral point resulting in a neutral point of second
order (Fig.\,\ref{hel6}). Here, the absolute value of
the monopole moment is highest\footnote{
Also that of the dipole moment. The fieldlines of that dipole
escape to the right if the second null point
is shifted from the right to the left of the origin, so that
the radial \lq outflow\rq~is stronger compared to the other
images. The displacement of the second neutral point influences the
inner boundary conditions for the field lines.}.
A completely different scenario is shown in Fig.\,\ref{hel7}. Here, 
we have only one null point off the $x$-axis.
This implies a complex circulation $\Gamma$, leading to a spiral structure
in the vicinity of the origin and emulating the Parker spiral.
This represents an inner boundary condition giving a strong azimuthal 
component (a strong winding)
of the magnetic field in the equatorial plane of the heliosphere.

\subsection{Discussion of the mirror symmetric case}

In the case of $\Gamma_{r}=0$ and $C_{1i}=0$, we get an equilibrium that
is mirror symmetric with respect to the $x$ axis. The magnetic null points read
\begin{equation}
  u=\frac{-\Gamma_{i}}{4\pi\, B_{S\infty}}\pm\sqrt{  \frac{C_{1r}}{B_{S\infty}}
+\frac{\Gamma_{i}^2}{16\pi^2\ B_{S\infty}^2}    }\, .
   \end{equation}
One can easily see that, if
\begin{equation}
\frac{C_{1r}}{B_{S\infty}}+\frac{\Gamma_{i}^2}{16\pi^2\ B_{S\infty}^2}<0
\quad\Longleftrightarrow\quad
C_{1r}<-\, \frac{\Gamma_{i}^2}{16\pi^2\ B_{S\infty}}\, ,
\end{equation}
two magnetic nulls occur, which are not lying on the $x$ axis,
and the astrosphere is open with respect to the counterflow direction.

For
\begin{equation}
C_{1r}=-\, \frac{\Gamma_{i}^2}{16\pi^2\ B_{S\infty}}\, ,
  \end{equation}
only one stagnation point exists. The second  null,
$u=0$, is not a real null point, but a pole.
The first stagnation point is far away from the origin (i.e. from the star).
The dipole part of the outflow, or the magnetic field, is like the resistance
of a flow around an obstacle.
Under the assumption $C_{1r}=B_{S\infty} R^2$, we get
\begin{equation}
u=x=\frac{-\Gamma_{i}}{4\pi\, B_{S\infty}}\pm
R\,\sqrt{ 1 + \frac{\Gamma_{i}^2}{16\pi^2\ B_{S\infty}^2 R^2}  } \, .
\end{equation}
For $\Gamma_{i}\rightarrow 0$, we get two stagnation points
($x=+R$ and $x=-R$), positioned
on a circle with radius $R$. Figure \ref{hel1} shows
that the fieldlines of this magnetic field are identical to the image of
streamlines of a flow around a body shaped like a circular
cylinder (keeping in mind the substitution
$\phi_{m}\rightarrow\phi$, $A\rightarrow\psi$, with ${\cal W}=\phi+i\psi$
as the hydrodynamical potential and $B_{S\infty}\rightarrow v_{\infty}$).
The circle is a separatrix that separates fieldlines
of different topology. This astrosphere, however, would have the disadvantage
of being a closed surface (line in 2D), and  there would be 
no possibility
that plasma could flow into the tail. This would imply a diffusion process, as
is described in Neutsch \& Fahr\,(\cite{Neutsch}).

\section{Pure sub- or super-Alfv\'enic flows}

A tangential discontinuity has to form due to magnetic shear,
Here two different, magnetized plasmas encounter.
At least, one has to expect a strong gradient perpendicular to the
magnetic field involving a non-singular current sheet. For typical
astrophysical plasmas, the structure of such a current sheet can be
derived by solving the coupled system of Vlasov
and Maxwell equations selfconsistently, assuming the symmetry of the previous
section. This kind of translation invariant plasma,
although collisionless, can be considered to
follow a quasi-Maxwellian
distribution function, so that we can use the solution of
Harris\,(\cite{Harris}).
Such a current sheet can be seen as the prototype of a current sheet
separating two plasmas.
To enable a constant asymptotic homogenous field  and to
mimic two astropause current
sheet positions, where the two symmetric branches of the potential field
separatrices should be localized,
we choose the following transformation equation
  \begin{eqnarray}
   \alpha (A) & = & \frac{\rm{sign}\left[\alpha' \right]\,A}{\sqrt{1-
M_{A\infty}^2}}
   + a_{1}\ln{\cosh{\frac{\displaystyle\frac{A}{B_{s\infty}} - y_{1}}{d_{1}}}}
   \nonumber\\
   & & + a_{2}\ln{\cosh{\frac{\displaystyle\frac{A}{B_{s\infty}}- 
y_{2}}{d_{2}}}}\,
,\label{altra}
   \end{eqnarray}
  with derivative
  \begin{eqnarray}
\alpha'(A)& = &\frac{\rm{sign}\left[\alpha' \right]}{\sqrt{1-M_{A\infty}^2}}
   + \frac{a_{1}}{B_{S\infty} d_{1}}
   \tanh{\frac{\displaystyle\frac{A}{B_{s\infty}}
   - y_{1}}{d_{1}}}
   \nonumber\\
   & & + \frac{a_{2}}{B_{S\infty}
d_{2}}\tanh{\frac{\displaystyle\frac{A}{B_{s\infty}}- y_{2}}{d_{2}}}
   \nonumber\\
     & = & \frac{B_{\infty}}{B_{S\infty}} + \frac{a_{1}}{B_{S\infty} d_{1}}
    \tanh{\frac{\displaystyle\frac{A}{B_{s\infty}}
    - y_{1}}{d_{1}}}
    \nonumber\\
    & & + \frac{a_{2}}{B_{S\infty}
d_{2}}\tanh{\frac{\displaystyle\frac{A}{B_{s\infty}}- y_{2}}{d_{2}}}\, ,
     \label{trafo1}
     \end{eqnarray}
   where ${\rm sign}[\alpha']$ indicates that the asymptotical magnetic
field can
   be parallel or anti-parallel to the asymptotical flow. Hence, with 
Eq.\,(\ref{trafo1}),
  \begin{equation}
   \rm{sign}[\alpha'] \equiv \rm{sign}[B_{\infty}].
    \end{equation}

   This transformation fulfills the symmetric asymptotic boundary conditions for
the magnetic field, given by
\begin{equation}
\lim\limits_{y\rightarrow\infty} \vec B =  B_{\infty}\vec e_{x} =
\lim\limits_{y\rightarrow - \infty}
\vec B\, ,\label{bound1}
  \end{equation}
where $B_{\infty}$ is a constant. In the case of an MHS equilibrium, where
  \begin{equation}
   A=B_{S\infty} y \qquad\textrm{with}\qquad B_{S\infty}>0\, , \label{equ3}
    \end{equation}
the second term and the third term of Eq.\,(\ref{altra}), which represent the outer
current sheets,
  are concentrated around $y_{1}$ and
  $y_{2}$, which should be the locations of the astropause envelope borders in 2D
(asymptotically).
  This implies that $\alpha'$ is the amplification factor of the static magnetic
field
  \begin{equation}
   \vec B= \vec\nabla\alpha\times\vec e_{z}=\alpha'\,\vec\nabla 
A\times\vec e_{z}=
   \alpha'\, B_{s\infty}\vec e_{x}\, ,
    \end{equation}
which shows the behaviour of the asymptotic magnetic field if the
magnetohydrostatic field is homogenous.
The stationary field at infinity ($x\rightarrow\infty$) can only depend on $y$
because $A$ converges to a value proportional to $y$ and, therefore,
$\alpha(A)\sim\alpha(y)$.
The symmetric boundary condition Eq.\,(\ref{bound1}) leads then to
\begin{equation}
  \frac{a_{1}}{d_{1}}=-\frac{a_{2}}{d_{2}}=:B_{1}\, ,\label{magnetic1}
  \end{equation}
which means that the current sheets have different signs 
(antiparallel currents).
The sign of the
Jacobian (here $\alpha'$) must be unique in the whole domain,
as there should be no roots
for a purely sub-Alfv\'enic or super-Alfv\'enic flow.
In addition, we make the assumption that $0<d_{1}=d_{2}\ll |y_{1}-y_{2}|$ and
$y_{1}>y_{2}$, $y_{1}=-y_{2}>0$
for the symmetric case (to ensure a highly symmetric equilibrium).

\subsection{Examples of noncanonical transformations for sub-Alfv\'enic flows}

With the above assumptions, Eqs.\,(\ref{trafo1}) and
Eq.\,(\ref{magnetic1}), we make certain that
\begin{equation}
\left|\frac{d\alpha}{dA}\right|=\left|\alpha'\right|>1 \label{magrestrict1},
   \end{equation}
so that the flow is sub-Alfv\'enic. With condition (\ref{magnetic1}),
Eq.\,(\ref{magrestrict1})
can be written as
  \begin{equation}
   \left|\alpha'(y=0)\right|=  \left|\frac{\rm{sign}\left[ \alpha' 
\right]}{\sqrt{1
- M_{A\infty}^2}} - \frac{2 B_{1}}{B_{S\infty}}\right| =
   \left|\frac{B_{\infty}}{B_{S\infty}} -\frac{2 
B_{1}}{B_{S\infty}}\right| > 1 \,
,\label{magrestrict2}
     \end{equation}
with
  \begin{equation}
   \frac{B_{\infty}}{B_{S\infty}}=\frac{\rm{sign}\left[ \alpha' 
\right]}{\sqrt{1 -
M_{A\infty}^2}}\, .
  \label{magrestrict3}
     \end{equation}
   This leads to the following restrictions
  \begin{eqnarray}
  \textrm{For}\quad B_{\infty} > 0 && \,\Leftrightarrow
\,\,\,\textrm{sign}[\alpha'] > 0 \nonumber\\
   \nonumber\\
  &&  2 B_{1} < B_{\infty}\left(1 - \sqrt{1 - M_{A\infty}^2}\,\right)\\
   \lor && 2 B_{1} > B_{\infty}\left(1 + \sqrt{1 - M_{A\infty}^2}\,\right)\, ,\\
  \nonumber\\
   \nonumber\\
  \textrm{for}\quad B_{\infty} < 0 && \,\Leftrightarrow
\,\,\,\textrm{sign}[\alpha'] < 0 \nonumber\\
  \nonumber\\
   && 2 B_{1} < B_{\infty}\left(1 + \sqrt{1 - M_{A\infty}^2}\right)\\
    \lor && 2 B_{1} > B_{\infty}\left(1 - \sqrt{1 - M_{A\infty}^2}\right)\, .
     \label{magrestrict4}
      \end{eqnarray}
%
%
%
The amplification factor $\alpha'$ of
the static asymptotic magnetic field $B_{s\infty}$ at the axis of 
symmetry tells us that  a magnetic jump occurs at the location of
the separatrix, because the magnetic fields from the inside and
outside converge to different values.
The strength of the jump is given by
  \begin{eqnarray}
    2 B_{1}=B_{\infty} - \alpha'(y=0) B_{S\infty} =
    B_{\infty}\left(1 - \frac{\sqrt{1 - M_{A\infty}^2}}{\sqrt{1 -
M_{A,i}^2}}\right)\, .
    \label{magjump}
    \end{eqnarray}
\begin{figure}
\resizebox{\hsize}{!}{\includegraphics{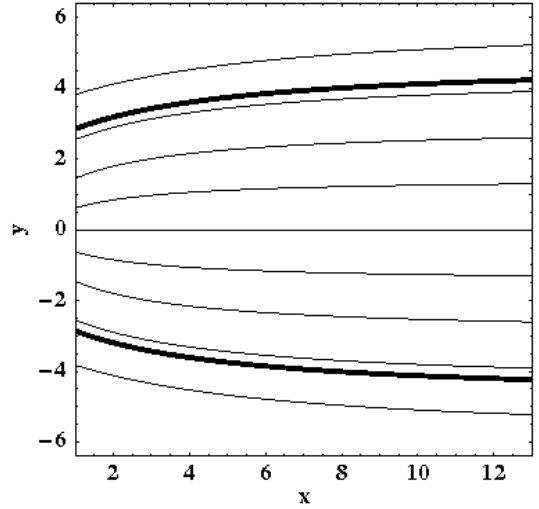}}
\caption{Regarding the field lines and especially the separatrix 
(thick line) one
can see that they are not identical with the isocontours of the 
electric current
plotted in Fig.\,\protect{\ref{currentlines1}}.}
\label{feldli1}
\end{figure}
\begin{figure}
\resizebox{\hsize}{!}{\includegraphics{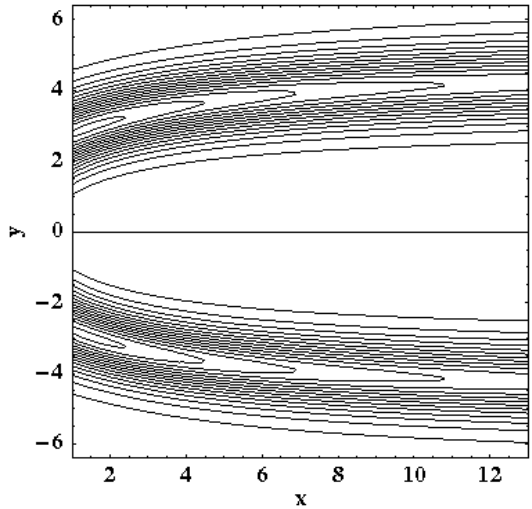}}
\caption{Isocontours of the current density, where the 
high current isocontours are obviously concentrated around the heliopause.
The curvature and closure of the current isocontours can
be seen clearly in contrast to the curved, but open separatrix line in
Fig.\,(\ref{feldli1}); only in the asymptotical 1D
region, i.e. for $x\rightarrow\infty$, the lines of
maximum current and the separatrix (astropause) shapes seem to converge.}
\label{currentlines1}
\end{figure}
Therefore, not all values of the inside magnetic field are allowed if the flow
is purely sub-Alfv\'enic. The inside magnetic field $B_{i}$ is given by
  \begin{equation}
   B_{\rm i}= B_{\infty} - 2 B_{1} =  B_{\infty}\, \frac{\sqrt{1 -
M_{A\infty}^2}}{\sqrt{1 - M_{A,i}^2}} \, .
    \label{magi1}
     \end{equation}
Therefore, the polarity cannot change from the outside to the inside.
With Eqs.\,(\ref{magi1}) and (\ref{magrestrict2}), we get a lower limit
for the inside magnetic field if the outside magnetic field
$B_{\infty}$ and the outside asymptotical Alfv\'en Mach number
$M_{A\infty}$ are given:
  \begin{eqnarray}
   \left| B_{\rm i}\right| > \sqrt{1 - M_{A\infty}^2}\,\left| 
B_{\infty}\right| =
B_{S\infty} \, .
    \label{magi2}
     \end{eqnarray}
Boundary conditions for the 1D case are given by the condition
Eq.\,(\ref{magi1}).
This implies that the outer and inner Alfv\'en Mach numbers
are the boundary conditions for Eq.\,({\ref{equ2}}),
together with the choice $B_{\infty}={\rm const}$. This determines
$B_{i}$, which cannot be prescribed as a boundary condition, since the problem
would be overdetermined.
With the above relations we see that the
possibility of setting boundary conditions is reduced,
due to the reflection symmetry condition for the magnetic field.
We choose the axis of symmetry in the
$x$-$y$ plane, the $x$-axis as second boundary, and demand only 
regularity on the other two boundaries
$x=1$ and $x=x_{\rm tail-end}$. Regularity is guaranteed by the behaviour of
the potential field and the transformation type in this domain.

We can also use the above transformation to fulfill the boundary 
condition for the asymptotical 1D region of a 2D field.
In this case, the boundary conditions are mapped
together with the mapping of the whole 2D potential field because we know in
advance that this will again be a stationary equilibrium state
with field-aligned incompressible flow.
Therefore, if for $x\rightarrow\infty$, the equilibrium converges
asymptotically to the 1D equilibria given by Eqs.\,(\ref{equ2}) and 
(\ref{trafo1}), this method
can be used. We can take any of the given potential fields in
the foregoing section, if we want to keep the potential character
of the magnetic field in the tail. This leads then to a 
2D sub-Alfv\'enic equilibrium state, writing
  \begin{eqnarray}
    \lim\limits_{x\rightarrow\infty}\alpha\left( A\left(x,y\right)\right) =
    \alpha(A_{\infty})\, ,\quad\textrm{with}\quad A_{\infty}= B_{S\infty} y\, .
     \end{eqnarray}
\begin{figure}
\resizebox{\hsize}{!}{\includegraphics{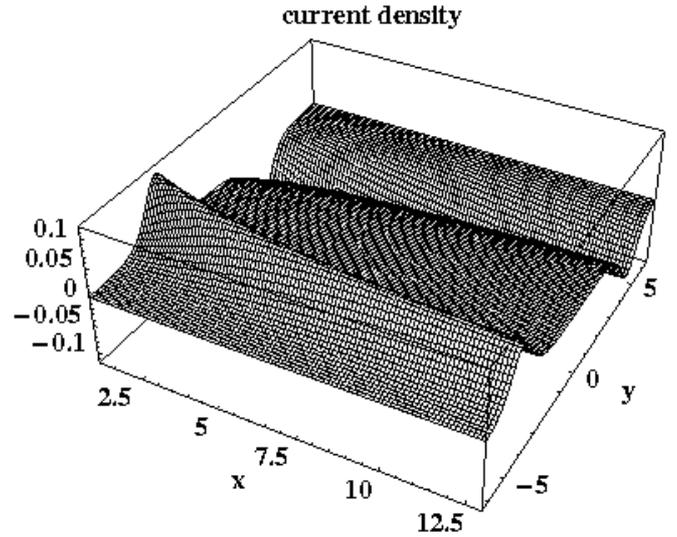}}
\caption{Shown is the current density and its increase towards the inner
astrosphere, i.e. the
downwind region of the reverse shock. Also visible are current sheets 
with a width of 100 AU.}
\label{current1}
\end{figure}

In Figs.\,\ref{feldli1} and \ref{currentlines1}, where we plotted 
bith the field lines
and separatrix and the isocontours of the current density, an
interesting feature of the
transformation can be seen: while in  magnetohydrostatics, where
$\vec B\cdot\vec\nabla P=0$ and $\vec j\cdot\vec\nabla P=0$ imply that the
current is constant on
  field lines, so that the current isocontours coincide with the
magnetic field lines,
the situation is now completely different.
  Comparing Fig.\,\ref{currentlines1} with the fieldlines in 
Fig.\,\ref{feldli1},
  it can be seen that, asymptotically, the field lines and isocontours of the
current density geometrically converge, but topologically they are 
different: while the field lines are open throughout the
  tail, the isolines of the current are closing in the vicinity of the 
separatrix
(i.e. the astropause).

To clearly show these differences in the field lines 
(Fig.\,\ref{feldli1}) and the
isocontours of the
current density (Fig.\,\ref{currentlines1}), a finite width of 
the current sheet
of 100\,AU has been used. These broad current sheets are visible in
Fig.\,\ref{current1} where
we have plotted the strength of the current density,
whose absolute value increases towards the termination shock region.

We now turn to the presentation of an example of a toy model that 
might represent the heliospheric tail region. First, we reduce the
width of the current sheet to a more realistic value.
That thickness can be estimated from
the fact that it should be larger than several ion gyroradii
(Neutsch \& Fahr\,\cite{Neutsch}; Fahr et al.\,\cite{Fahr5}).
The ion gyroradius is at least of the order of 
$10^2$--$10^3$\,km,  and we set the
width of the current sheet to 10\,AU.

\begin{figure}
\resizebox{\hsize}{!}{\includegraphics{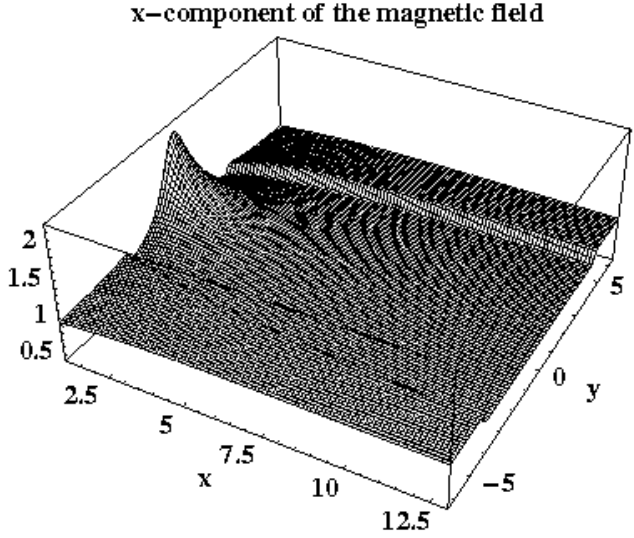}}
\caption{The $x$-component of the magnetic field. It rises towards the location
of the termination shock caused by the magnetic monopole.
Clearly visible are the jumps at the locations of the current sheets.}
\label{BX2}
\end{figure}

\begin{figure}
\resizebox{\hsize}{!}{\includegraphics{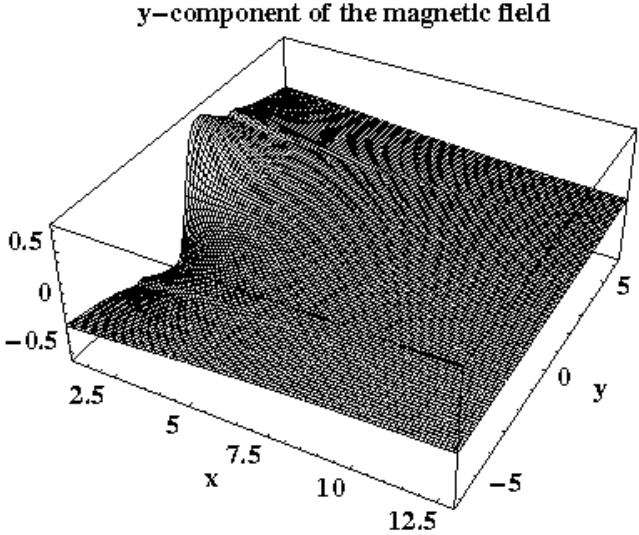}}
\caption{The $y$-component of the magnetic field also shows a strong 
increase in its absolute value towards the termination shock region.}
\label{BY2}
\end{figure}

In addition, we can fall back on
the measurements and estimated values given in Frisch et al.\,(\cite{Frischb})
(Table 1, Model 2)
\footnote{The extreme value of the magnetic field
is considered too high
by P. Frisch, but preferred by Cox \& Helenius\,(\cite{Cox}).}.
For the outside magnetic field we take
$B_{\infty}=5\,\mu$G, for the proton density
$n_{i}\approx n_{e}=0.1$\,cm$^{-3}$ and for the velocity
${\rm v}_{\infty}=25$\,km\,s$^{-1}$,
so we obtain an interstellar Alfv\'en Mach number
$M_{A\infty}\approx 0.72$. Assuming an inner Alfv\'en
Mach number of about 0.52, the inner magnetic field in the vicinity 
of the $x$-axis
becomes about $4\,\mu$G.
Taking the relation for the inner magnetic field, Eq.\,(\ref{magi1}) and
$d_{1}=10$\,AU, we are able
to calculate the transformation, using
Eqs.\,(\ref{trafo1}) and (\ref{magnetic1}):
\begin{eqnarray}
\alpha'(A) &=& \frac{1}{\sqrt{1-M_{A\infty}^2}} \left(\frac{B_{1}}{B_{\infty}}
     \tanh{\frac{\displaystyle\frac{A}{\sqrt{1 - M_{A\infty}^2}\, B_{\infty}}
    - y_{1}}{d_{1}}} \right.
    \nonumber\\
    & & \left. -\frac{B_{1}}{B_{\infty}}
\tanh{\frac{\displaystyle\frac{A}{\sqrt{1 - M_{A\infty}^2}\, B_{\infty}}+
y_{1}}{d_{1}}}\right)\, ,
     \label{trafoexample}
\end{eqnarray}
where for $B_{1}$ we used the definition given by Eq.\,(\ref{magjump}):
\begin{eqnarray}
    B_{1}=\frac{1}{2}\,\left(B_{\infty} - \alpha'(y=0) B_{S\infty}\right) =
    \frac{B_{\infty}}{2}\left(1 - \frac{\sqrt{1 - M_{A\infty}^2}}{\sqrt{1 -
M_{A,i}^2}}\right) \, .
    \label{magneticjump}
    \end{eqnarray}
\begin{figure}
\resizebox{\hsize}{!}{\includegraphics{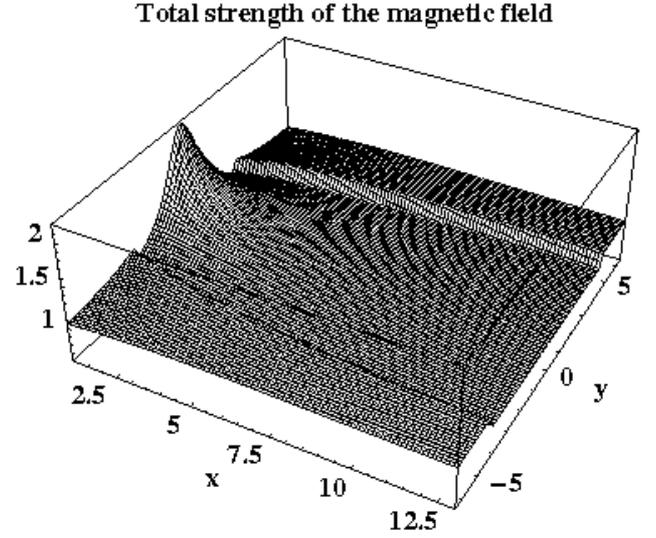}}
\caption{Total strength of the magnetic field. The dominant 
contribution is from
the $x$-component while,
the contribution of the $y$-component is almost negligible.}
\label{Bfeld2}
\end{figure}


\begin{figure}
\resizebox{\hsize}{!}{\includegraphics{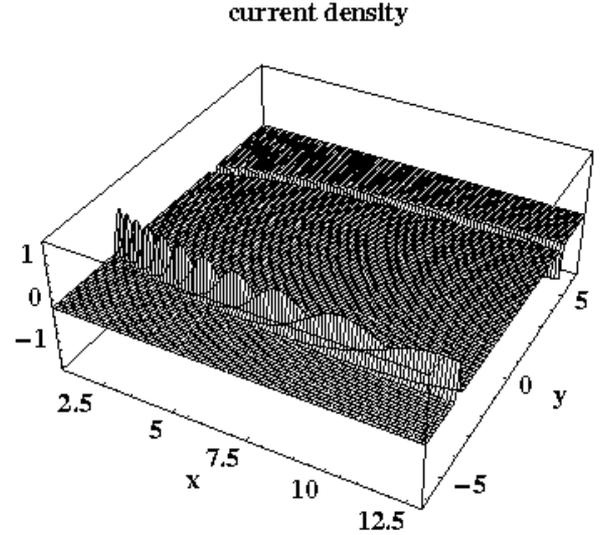}}
\caption{The strength of the current density increases 
steeply towards the current sheets that have
a width of 10\,AU. The wavy shape of the peak current 
density is not physical but an artefact due to numerics.
The current is normalized to units of $2.65\times 10^{-17}$\,Am$^{-2}$.}
\label{current2}
\end{figure}

\begin{figure}
\resizebox{\hsize}{!}{\includegraphics{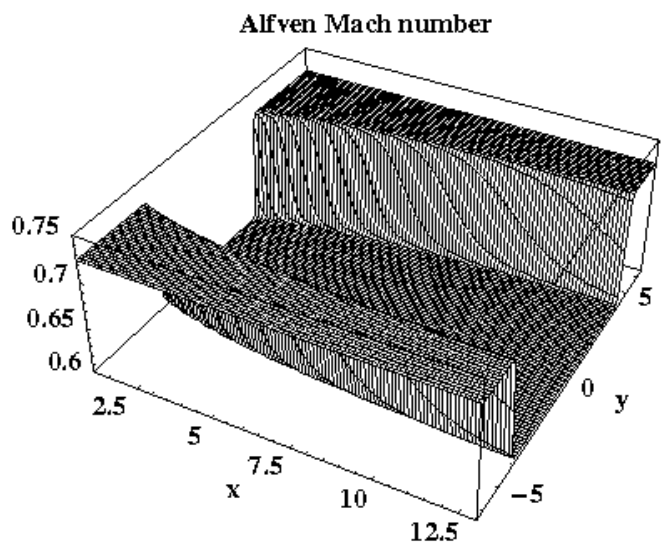}}
\caption{This plot shows the Mach number and its strong gradient across the
heliopause.}
\label{machnumber2}
\end{figure}

The results for the different parameters are shown in 
Figs.\,\ref{BX2}--\ref{machnumber2},
in Figs.\,\ref{BX2} and \ref{BY2} we plotted the $x$- and $y$-components of the
magnetic field in
units of $5\,\mu$G. Towards the termination shock, the $x$-component grows
especially around the $x$-axis
because it is approaching the mapped magnetic monopole. The jumps due to the
current sheet are
of the order $1\,\mu$G. The $y$-component of the magnetic field is 
not symmetric with respect to the $x$-axis. Instead,
a gradient arises due to the monopole. The contribution of
the $y$-component to the total magnetic field strength is, however, 
very small, as can be seen in Fig.\,\ref{Bfeld2}(compared with
Fig.\,\ref{BY2}).
Figure\,\ref{current2} shows the strength of the current density. It is only
different from zero
at the locations of the current sheets where it shows steep gradients.
In the last figure (Fig.\,\ref{machnumber2}) we display the behaviour of the
Alfv\'en Mach number
in the tail. It shows a strong gradient at the locations of the current sheets.
This gradient
even increases towards the termination shock.

%
%

%

\section{Discussion and conclusions}

We present a method for calculating nonlinear MHD equilibria
with an incompressible field-aligned flow. This method is applied
to the scenario of a flow of interstellar plasma around the 
plasma bubble of a strong magnetized stellar wind. We use the classical 
method of conformal mapping of flows
around an obstacle as the starting point of our calculations.

We exclude violent structures (shocks) of flows in order to 
concentrate on the study of (i)
the
geometry of the contact surface, (ii) the surface currents that are coupled to
the inner and outer magnetic fields and the Alfv\'en Mach number as
boundary conditions. The advantage of such a method is its high 
flexibility in modelling the tail of
stellar winds and the surrounding interstellar medium wind. What we 
need at least is information on the singular points (stagnation- and
magnetic neutral points), their numbers, and their orders. Hence, it would be
better if in situ measurement of the magnetic field structure could be made.
Within the next decade, this is only possible for our own astrosphere, the
heliosphere.

In this paper, we restricted ourselves to thin nonsingular current sheets
that have the special shape of a Harris-sheet or z-pinch configuration. The
validity can only be justified within a multi-fluid theory or, 
better, within the
framework of kinetic plasma theory and a detailed knowledge of the plasma
environment in astrotails. Again, this aspect will be studied best
observationally (in the next decade) for our heliosphere.

As an improvement, additional current sheet structures should be taken into
account, as e.g. the heliospheric current sheet is believed to extend beyond
the heliospheric termination shock (see e.g. Pogorelov et al.,
\cite{Pogorelov}).

Our future aims are to use nonlinear static MHD equilibria as original
equilibria, where the correlation between magnetic neutral points will be much
more complex. In addition, we have to find mappings with corresponding
boundary conditions.
Another important point is the symmetry we have taken
into account: symmetry of such configurations can
be broken easily by an angle between the magnetic field and the 
probable flow direction in the vicinity of the
heliosphere (see e.g. Frisch\,\cite{Frisch}). Such symmetry breaking will
probably exclude an axially symmetric treatment of the problem.
Nevertheless axially symmetric static equilibria can, in priciple, 
be used. In the case of pole-on counterflows of
magnetized stellar winds, axial symmetry should be applied.

For further investigation, transformations should be used that allow for
transitions from sub- to super-Alfv\'enic flows perpendicular to the magnetic
field lines, as briefly described in Gebhardt \& Kiessling\,(\cite{Geki}).
Their application is much more complicated, as it is necessary to 
introduce four
Euler potentials for the representation of the velocity and the magnetic field.

\begin{acknowledgements}
This project was partly
supported by the grant of the project No. 98030 of
the PECS programme. D.H.N. thanks M. Karlicky for helpful discussions. 
   \end{acknowledgements}

\Online

\begin{appendix}

\section{Derivation of the transformation equations}

All solenoidal fields (i.e. vector fields with
vanishing divergence) can be described locally by means of
two scalar functions $f$ and $g$

\begin{equation}
\vec B=\vec\nabla f\times\vec\nabla g\, .
\end{equation}
Applying the scalar product $\vec\nabla f$ or $\vec\nabla g$,
it follows that

\begin{equation}
  \vec B\cdot\vec\nabla f=0\, ,~\quad\textrm{and}\quad~\vec 
B\cdot\vec\nabla g=0\,
.
   \end{equation}

Since the potentials $f$ and $g$ are constant on field lines,
besides the pressure in magnetohydrostatic equilibria $P_{S}$,
we can understand $P$ as a function of $f$ and $g$
(i.e. intersections of $f=\textrm{const}$ and $g=\textrm{const}$
are field lines).
We now want to consider the magnetohydrostatic equations with
$\vec B_{S}=\vec\nabla f\times\vec\nabla g$, where
$\vec B_{S}$ is the magnetic field of a known static MHD equilibrium. With
the aforementioned equations, it follows that
\begin{eqnarray}
\vec\nabla P_{S} &=& \frac{1}{\mu_{0}}\,(\vec\nabla\times\vec B_{})\times\vec
B\, ,\\
\nonumber\\ \textrm{and with}~\, ~\vec\nabla P_{S} &=& \frac{\partial
   P_{S}}{\partial f}\,\vec\nabla f+\frac{\partial P_{S}}{\partial 
g}\,\vec\nabla g\, ,
~
\end{eqnarray}
we can extract the equations of motion
\begin{eqnarray}
    \frac{\partial P_{S}}{\partial f}&=&\vec\nabla g\cdot
      \vec\nabla\times(\vec\nabla f\times\vec\nabla g)\, ,\\ \nonumber\\
     - \frac{\partial
       P_{S}}{\partial g}&=&\vec\nabla f\cdot\vec\nabla\times
       \left(\vec\nabla f\times\vec\nabla g\right)\, .
   \end{eqnarray}
The scalar potentials are often called {\it Euler potentials}.
%
%
%
Using unmatched Euler potentials $\alpha$ and $\beta$,
with $f=f(\alpha,\beta)$ and $g=g(\alpha,\beta)$, it follows from
adopting the functional determinant or Poisson brackets defined as
  %
\begin{equation}
[f,g]_{\alpha,\beta}=\frac{\partial f}{\partial\alpha}\frac{\partial
g}{\partial\beta}
-\frac{\partial f}{\partial\beta}\frac{\partial g}{\partial\alpha}\, ,
\end{equation}
that the following relation holds:
\begin{equation}
\vec B_{S}=
\vec\nabla f\times\vec\nabla g=[f,g]_{\alpha,\beta}\,\vec\nabla\alpha\times
\vec\nabla\beta:=[f,g]_{\alpha,\beta}\,\vec B\, .
\end{equation}
Therefore,
\begin{eqnarray}
\mu_{0}\,\vec\nabla P_{S} & = & \left(\vec\nabla\times\left(\vec\nabla
   f\times\vec\nabla g\right)\right)\times
     \left(\vec\nabla f\times\vec\nabla g\right)\nonumber\\
  & = & \left(\vec\nabla\times ~
   [f,g]_{\alpha,\beta}\,\vec\nabla\alpha\times\vec\nabla\beta\right)\times\,
   [f,g]_{\alpha,\beta}\,\vec\nabla\alpha\times\vec\nabla\beta\nonumber\\
  & = & \left( \vec\nabla
   [f,g]_{\alpha,\beta}\times\,\vec\nabla\alpha\times\vec\nabla\beta
\right)\times\,
[f,g]_{\alpha,\beta}\left(\vec\nabla\alpha\times\vec\nabla\beta\right)\nonumber\\
  & & \, +\left(
[f,g]_{\alpha,\beta}\vec\nabla\times\left(\vec\nabla\alpha\times\vec\nabla\beta 
\right)  \right)\times
[f,g]_{\alpha,\beta}\left(\vec\nabla\alpha\times\vec\nabla\beta\right)\nonumber\\
  & = &
\left(\frac{1}{2}\vec\nabla[f,g]_{\alpha,\beta}^2\times\left(\vec\nabla\alpha
\times\vec\nabla\beta\right)
\right)\times\left(\vec\nabla\alpha\times\vec\nabla\beta\right)\nonumber\\
  & &
\,+[f,g]_{\alpha,\beta}^2\,\vec\nabla\times\left(\vec\nabla\alpha\times
\vec\nabla\beta\right)\times\left(
\vec\nabla\alpha\times\vec\nabla\beta\right)\nonumber\\
  & =&
[f,g]_{\alpha,\beta}^2\,\vec\nabla\times\left(\vec\nabla\alpha\times\vec\nabla
\beta\right)\times\left(\vec\nabla\alpha\times\vec\nabla\beta\right)\nonumber\\
  & & ~\,+\underbrace{\left(\frac{1}{2}\,\vec\nabla
  [f,g]_{\alpha,\beta}^2\cdot\left(\vec\nabla\alpha\times\vec\nabla\beta\right)
\right)}_{\equiv 0}\left(\vec\nabla\alpha\times\vec\nabla\beta\right)\nonumber\\
  & & ~\,-\frac{1}{2}\,\left(\vec\nabla\alpha\times\vec\nabla\beta\right)^2
\vec\nabla [f,g]_{\alpha,\beta}^2\nonumber\\
  & = &
[f,g]_{\alpha,\beta}^2\,\vec\nabla\times\left(\vec\nabla\alpha\times\vec\nabla
\beta\right)\times\left(\vec\nabla\alpha\times\vec\nabla\beta\right)\nonumber\\
  & & ~\,-\frac{1}{2}\left(\vec\nabla\alpha\times\vec\nabla\beta\right)^2
\vec\nabla [f,g]_{\alpha,\beta}^2\label{ham1}
\end{eqnarray}
If we identify the equilibrium magnetic field with
\begin{equation}
\vec\nabla\alpha\times\vec\nabla\beta\equiv\vec B\, ,\label{transform1}
\end{equation}
the sum of thermal and ram pressure with
\begin{equation}
P_{S}\equiv\Pi\equiv P+\rho|\vec{\rm v}|^2/2\, ,\label{transform2}
\end{equation}
and the corresponding equilibrium current density of the {\it stationary}
equilibrium with
\begin{equation}
\vec\nabla\times(\vec\nabla\alpha
\times\vec\nabla\beta)\equiv\mu_{0}\vec j\, ,\label{transform3}
\end{equation}
we find
\begin{equation}
\vec\nabla\Pi=\frac{1}{\mu_{0}}[f,g]_{\alpha,\beta}^2\,(\vec j\times\vec B)
  -\frac{1}{2\mu_{0}}\vec B^2\,\vec\nabla [f,g]_{\alpha,\beta}^2\, .
\label{ham2}
\end{equation}

We recognize the identical form of the last term on the right side
of Eq.\,(\ref{ham2}) with the right-hand side of the equation of motion
(Eq.\,\ref{bewegl3b}), which is

\begin{equation}
\vec\nabla\Pi=\frac{1}{\mu_{0}}(1-M_{A}^2)(\vec j\times\vec B)-
     \frac{1}{2\mu_{0}}\vec B^2\,\vec\nabla(1- M_{A}^2)
\label{bewegl3b}\, ,
\end{equation}
where we identify the Poisson brackets of Eq.\,(\ref{ham2}) with
the Alfv\'en Mach numer dependent expression in Eq.\,(\ref{bewegl3b}).
Then it follows that

\begin{equation}
1-M_{A}^2\equiv [f,g]_{\alpha,\beta}^2>0\, .
  \end{equation}

For non-canonical transformations, the Poisson brackets has a
non-constant value or a value different from unity. For
canonical transformations, the Mach number is zero, or for a constant
and non-unity value of the Poisson brackets, the Mach number turns out
to be constant. It is therefore possible to map known solutions of the
magnetohydrostatic equations with the help of non-canonical
transformations into stationary solutions with a sub-Alfv\'enic flow.
For $M_{A}>1$, we find an analogy with the equations of incompressible
stationary hydrodynamics (see Gebhardt \& Kiessling\,\cite{Geki}).
Application of the same transformation as in the case of
the sub-Alfv\'enic equilibrium results in the same magnetic field and therefore
in the same electric current. On the
other hand, for Mach numbers with $M_{A}^2\geq 2$, we cannot find any 
sub-Alfv\'enic Mach number or solution.


The pressure is going to be inverted with respect to the 
sub-Alfv\'enic pressure to become:
\begin{eqnarray}
  \Pi_\textrm{super-Alfv\'enic}&=&\Pi_{H}-\Pi_\textrm{sub-Alfv\'enic}\, .
\nonumber\\
  \end{eqnarray}
Here, $\Pi_{H}$ is a background pressure, which guarantees that the
thermal pressure stays positive everywhere and fullfills the physical
conditions (as e.g. $v_{S} > v_{A}$), as well as the boundary conditions.


As shown above, it is possible to map known solutions of the
magnetohydrostatic equations via the non-canonical transformations
into stationary solutions with sub-Alfv\'enic flow. If there exists
a non-canonical transformation $f=f(\alpha,\beta)$ and $g=g(\alpha,\beta)$
or $\alpha=\alpha(f,g)$ and $\beta=\beta(f,g)$, then those stationary
fields are given by


\begin{eqnarray}
   \vec B &=&\vec\nabla\alpha\times\vec\nabla\beta=
           \frac{1}{[f,g]_{\alpha,\beta}}\,\vec\nabla f\times\vec\nabla g=
            \frac{1}{\sqrt{1-M_{A}^2}}\,\vec B_{S}\, ,\nonumber\\
   \vec v 
&=&\frac{M_{A}\left(\alpha\left(f,g\right),\beta\left(f,g\right)\right)}{
\sqrt{\mu_{0}\rho\left(\alpha\left(f,g\right),\beta\left(f,g\right)\right)}}\vec 
B
         =\frac{M_{A}(f,g)}{\sqrt{\mu_{0}\rho(f,g)}}\vec B \nonumber\\
          &=& \frac{M_{A}}{\sqrt{(1-M_{A}^2)\,\mu_{0}\rho}}\,\vec B_{S}\,
,\nonumber\\
         P&=&\Pi(f,g)-\frac{M_{A}^2}{2}|\vec B|^2
           = P_{S}(f,g)-\frac{M_{A}^2}{2\sqrt{1-M_{A}^2}}|\vec B_{S}|^2\,
,\nonumber\\
     \rho &=& \rho(f,g)\, .\label{trafo2}
          \end{eqnarray}

\section{Trajectories}

If $f$ is a flux function, then $f$ is constant on field lines.
For $\Delta A=-\mu_{0}\, dP/dA=J(A)$ and with the help of the
implicit function given by $\chi=\chi(x,y)$, where $\chi_{0}=\chi(x,y)
=\textrm{const}$, it follows that $A(\chi)=\textrm{const}$.
The equation $\chi=\chi(x,y)=\textrm{const}$ therefore describes 
the bundle of the field lines of $A(\chi)$.


The equation
\begin{equation}
\Delta A=\frac{d^2A}{d\chi^2}(\vec\nabla\chi)^2 +
          \frac{dA}{d\chi}\,\Delta\chi=-\mu_{0}\, 
dP/dA=J\left(A\left(\chi\right)
             \right)
\end{equation}
can therefore be regarded as the differential equation for $A$
as a function of $\chi$, if we have
\begin{equation}
  \frac{\partial\left(
      \left|\vec\nabla\chi\right|^{-2}\Delta\chi,\chi\right)}{\partial(x,y)}=0
  \label{pigra}
   \end{equation}
in the case of a vanishing current function (i.e. a potential field), i.e. for
$J(A)=0$. This results in a non-linear partial differential equation.


\subsection{Radial magnetic fields}

For $\Delta A=-\mu_{0}\, dP/dA=J(A)=0$ and for radial trajectories,
it follows that, for $y/x=\textrm{const}$ on straight lines, the function
$A(\chi) = \textrm{const}$. $\chi$ is thereby regarded as a function
of $x$ and $y$, with $\chi(x,y):= y/x$.
Application of the Laplace operator on $A(\chi (x,y))$ results in
\begin{eqnarray}\label{DeltaA}
\Delta A & = & 
A''(\chi)(\frac{y^2}{x^4}+\frac{1}{x^2})+A'(\chi)\,\frac{2y}{x^3}=0
\nonumber\\
    & & \, \Rightarrow\, ~ A''(\chi)(\chi^2+1)+2A'(\chi)\chi =0\, .
\end{eqnarray}
The second row of Eq.\,(\ref{DeltaA}) can be expressed as
\begin{equation}
\frac{d}{d\chi}\, \left(A'\left(\chi\right)\left(1+\chi^2\right)\right) = 0\, .
\end{equation}
Integration leads to
  \begin{equation}
  A(\chi)=\textrm{const}\cdot\arctan(\chi)+A_{0}\, ,
  \end{equation}
with $A_{0}$ as a constant of integration. The function $A(\chi)$ can be
expressed in the framework of a 2D multipole expansion with the help of the
imaginary part of the line current, i.e. with the radial
magnetic field (represented by the imaginary part of the complex logarithm).

\end{appendix}

\end{document}